\documentclass[runningheads]{llncs}
\usepackage[T1]{fontenc}
\usepackage{graphicx}

\usepackage{xcolor}
\usepackage{amsmath, amssymb}
\usepackage[ruled, noend, linesnumbered]{algorithm2e}
\usepackage{bm}
\usepackage{stmaryrd}
\usepackage{tikz}
\usepackage{tcolorbox}
\tcbuselibrary{breakable}
\tcbuselibrary{skins}
\usepackage{mathtools}
\usepackage{rotating}
\usepackage{adjustbox}
\usepackage{tabularray}
\usepackage{hhline}
\usepackage{multirow}
\usepackage{cite}
\usepackage{paralist}
\usepackage{makecell}
\usepackage{hyperref}
\usepackage{cleveref}
\usepackage{wrapfig}
\usepackage[firstpage]{draftwatermark}
\usepackage{orcidlink}

\newif\iflong
\longfalse

\newif\ifextended
\extendedtrue

\newif\ifrange
\rangetrue

\newcommand{\powerset}[1]{{\mathcal{P}(#1)}}
\newcommand{\perm}[1]{\mathfrak{S}_{#1}}

\newcommand{\weakenop}{\mathcal{W}}
\newcommand{\weaken}[3]{{\mathcal{W}_{#1}\left({#2},{#3}\right)}}
\newcommand{\up}[1]{{\uparrow}{#1}}

\newcommand{\upclose}[1]{{\uparrow}{#1}}

\newcommand{\init}{\iota}
\newcommand{\tr}{\tau}

\newcommand{\trans}{\mathcal{T}}
\newcommand{\abs}[1]{{#1}^\sharp}
\newcommand{\lfp}{\text{lfp}}

\newcommand{\state}{s}
\newcommand{\signature}{\Sigma}
\newcommand{\vars}{V}
\newcommand{\struct}{\sigma}
\newcommand{\univ}{\mathcal{U}}
\newcommand{\interp}{\mathcal{I}}
\newcommand{\asgn}{\mu}
\newcommand{\asgnb}{\nu}
\newcommand{\structs}[1]{\mathbf{structs}[{#1}]}
\newcommand{\dom}[1]{\mathbf{dom}({#1})}
\newcommand{\stateupdate}[2]{{#1}\overleftarrow{\cup}{#2}}

\newcommand{\doublesig}{\signature\cup\signature'}

\newcommand{\states}{\mathbb{S}}
\newcommand{\join}{\sqcup}
\newcommand{\cdomain}{\mathcal{C}}
\newcommand{\adomain}{\mathcal{A}}
\newcommand{\csqsubseteq}{\sqsubseteq_\cdomain}
\newcommand{\asqsubseteq}{\sqsubseteq_\adomain}

\newcommand{\tup}[1]{\Bar{#1}}
\newcommand{\seq}[1]{\langle{#1}\rangle}
\newcommand{\tupphi}{\tup{\phi}}
\newcommand{\tuppsi}{\tup{\psi}}
\newcommand{\emptytup}{\epsilon}
\newcommand{\indices}[1]{[{#1}]}
\newcommand{\ordered}[2]{{{\seq{#2}}_{#1}}}
\newcommand{\card}[1]{{|{#1}|}}

\newcommand{\orA}[2]{{{\bm\vee}_{#1}[{#2}]}}
\newcommand{\orAB}[2]{{{\bm\vee}[{#1}, {#2}]}}
\newcommand{\andA}[2]{{{\bm\wedge}_{#1}[{#2}]}}
\newcommand{\andAB}[2]{{{\bm\wedge}[{#1}, {#2}]}}
\newcommand{\forallA}[2]{{{\bm\forall}_{#1}[{#2}]}}
\newcommand{\existsA}[2]{{{\bm\exists}_{#1}[{#2}]}}
\newcommand{\existsforall}{{\mathrlap{\bm\exists}\hspace{0.31em}{\bm\forall}}}
\newcommand{\existsforallA}[2]{{\existsforall_{#1}[{#2}]}}

\newcommand{\rightmodels}{\Relbar\joinrel\mathrel{|}}

\newcommand{\lang}{\mathcal{L}}
\newcommand{\aset}{A}
\newcommand{\langa}{\lang_{\aset}}
\newcommand{\clang}[1]{{\bm{#1}}}
\newcommand{\stdset}[1]{\text{Set}[{#1}]}
\newcommand{\stdmap}[2]{\text{Map}[{#1}, {#2}]}
\newcommand{\lset}[1]{\text{LSet}[{#1}]}
\newcommand{\tnode}{\text{TreeNode}}
\newcommand{\unsat}[2]{{#1}|_{\not\rightmodels {#2}}}
\newcommand{\subsuming}[2]{{#1}|_{\sqsubseteq {#2}}}

\newcommand{\orandAB}[2]{{\circ[{#1}, {#2}]}}
\newcommand{\quantA}[2]{{Q_{#1}[{#2}]}}

\newcommand{\quant}{Q}
\newcommand{\veewedge}{\circ}

\renewcommand{\phi}{\varphi}

\newcommand{\flyvy}{Flyvy}

\allowdisplaybreaks

\Crefname{section}{Sec.}{Secs.}
\Crefname{algorithm}{Alg.}{Algs.}
\Crefname{definition}{Def.}{Defs.}
\Crefname{figure}{Fig.}{Figs.}
\Crefname{example}{Ex.}{Exs.}
\Crefname{appendix}{App.}{Apps.}
\Crefname{theorem}{Thm.}{Thms.}
\Crefname{corollary}{Cor.}{Cors.}

\begin{document}
\title{Efficient Implementation of an Abstract Domain of Quantified First-Order Formulas}
\author{Eden Frenkel\inst{1}\orcidlink{0009-0009-4589-2173} \and
Tej Chajed\inst{2}\orcidlink{0000-0002-9889-4828} \and
Oded Padon\inst{3}\orcidlink{0009-0006-4209-1635} \and
Sharon Shoham\inst{1}\orcidlink{0000-0002-7226-3526}}

\authorrunning{E. Frenkel et al.}

\institute{Tel Aviv University, Tel Aviv, Israel \and
University of Wisconsin-Madison, Madison, WI, USA \and
VMware Research, Palo Alto, CA, USA}

\maketitle

\SetWatermarkAngle{0}
\SetWatermarkText{\raisebox{12.6cm}{%
\hspace{0.1cm}%
\href{https://zenodo.org/doi/10.5281/zenodo.10938367}{\includegraphics{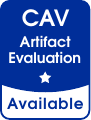}}%
\hspace{9cm}%
\includegraphics{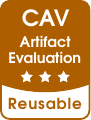}%
}}

\begin{abstract}
  This paper lays a practical foundation for using abstract interpretation with an abstract domain that consists of sets of quantified first-order logic formulas. This abstract domain seems infeasible at first sight due to the complexity of the formulas involved and the enormous size of sets of formulas (abstract elements). We introduce an efficient representation of abstract elements, which eliminates redundancies based on a novel syntactic subsumption relation that under-approximates semantic entailment. We develop algorithms and data structures to efficiently compute the join of an abstract element with the abstraction of a concrete state, operating on the representation of abstract elements.
To demonstrate feasibility of the domain, we use our data structures and algorithms to implement a symbolic abstraction algorithm that computes the least fixpoint of the best abstract transformer of a transition system, which corresponds to the strongest inductive invariant. 
We succeed at finding, for example, the least fixpoint for Paxos (which in our representation has 1,438 formulas with $\forall^*\exists^*\forall^*$ quantification) in time comparable to state-of-the-art property-directed approaches.

\end{abstract}

\setcounter{footnote}{0}

\section{Introduction}

Recent years have seen significant progress in automated verification based on first-order logic.
In particular, quantified first-order formulas have been used to model many
systems, their properties and their inductive invariants~\cite{paxos-made-epr,%
mcmillan-decidable-ivy,%
linked-lists-epr,%
modularity-for-decidability,%
verification-modulo-axioms,%
bounded-horizon,%
ic3po,%
natural-proofs,%
fossil,%
distai,duoai,%
updr,swiss,%
ic3po-nfm,%
i4,p-fol-ic3,%
induction-duality,%
vericon-networks%
}.
Automatic verification in this domain is challenging because of the combination of the complexity of first-order reasoning performed by solvers and the enormous search space of formulas, especially due to the use of quantifiers.
Despite these challenges, there are impressive success stories
of automatically inferring quantified inductive invariants for complex distributed and concurrent algorithms~\cite{ic3po,distai,duoai,%
updr,swiss,%
ic3po-nfm,%
i4,p-fol-ic3,%
induction-duality%
}.

Previous works on invariant inference for first-order logic search for invariants in the form of sets of formulas (interpreted conjunctively) from some language of quantified first-order formulas.
Each approach fixes some restricted, typically finite (but extremely large) language $\lang$,
and searches for a set of $\lang$-formulas that form an inductive invariant
using sophisticated heuristics and algorithmic techniques, such as
property-directed reachability (IC3)~\cite{updr,p-fol-ic3},
incremental induction~\cite{swiss,induction-duality},
generalization from finite instances~\cite{i4,ic3po-nfm},
and clever forms of pruning and exploration~\cite{distai,duoai}.
While prior techniques can successfully handle some challenging examples, the accumulation of specially-tailored techniques makes the results computed by these techniques unpredictable, and makes it hard to extend or improve them.

Abstract interpretation~\cite{cc77,cc79} suggests a more systematic approach for the development of verification algorithms based on logical languages, where we consider
sets of $\lang$-formulas as elements in an abstract domain.
The abstraction %
of a set of states $S$ in this domain is given by $\alpha(S) = \{ \phi \in \lang \mid \forall s\in S.\ s\models \phi\}$, i.e., the formulas that are satisfied by all states in the set.
Algorithms based on abstract interpretation are better understood and are easier to combine, extend, and improve. However, an abstract domain of quantified first-order formulas seems infeasible: for interesting systems, the abstract elements involved in proofs would contain an astronomical number of formulas.

The main contribution of this work is %
to develop algorithms and data structures that
make an abstract domain based on quantified first-order formulas feasible.
Working with this abstract domain introduces two main challenges:
(i)~efficiently storing and manipulating abstract elements comprising of many formulas,
and (ii)~overcoming solver limitations when reasoning over them.
This work focuses on the first challenge and adopts ideas from prior work~\cite{p-fol-ic3} to deal with the second.
Our techniques lay a practical foundation for using an abstract interpretation approach to develop new analyses in the domain of quantified first-order formulas.
We demonstrate feasibility of the abstract domain by applying it to an analysis of several intricate distributed protocols.

Our first key idea is to design a \emph{subsumption relation} for quantified first-order formulas and use it to
represent abstract elements (sets of formulas) more compactly, pruning away some formulas that are redundant since they are
equivalent to or are entailed by another formula.
Subsumption over propositional clauses (disjunctions of literals) is traditionally used for similar pruning purposes
(e.g.,~\cite{mcmillan-k-clause}),
but the generalization to first-order formulas, which include disjunction, conjunction, and quantification, is novel.

The second key ingredient of our approach is a way to manipulate abstract elements in our representation.
Rather than implementing the standard operations of $\alpha$ (abstraction) and $\join$ (abstract join),
we observe that our subsumption-based representation makes it more natural to directly implement
an operation that computes the join of an abstract element $a$ with the abstraction of a given concrete state $\state$, i.e., $a \join \alpha(\{\state\})$.
This operation can be used to compute the abstraction of a set of states,
and can also be used to compute the least fixpoint of the best abstract transformer (in the style of symbolic abstraction~\cite{symabs2004}).
The crux of computing $a \join \alpha(\{\state\})$ is to \emph{weaken} the formulas in the representation of $a$ to formulas that are subsumed by them and that $\state$ satisfies.

Finally, the third key ingredient of our approach is a data structure for storing a set of formulas,
with efficient filters for (i)~formulas that a given state does not satisfy, and (ii)~formulas that subsume a given formula. This data structure is then used to store abstract elements, and the filters make the implementation of $a \join \alpha(\{\state\})$ more efficient.

While the paper presents the ingredients of our approach (subsumption, weakening, and the data structure) sequentially, they are interconnected; they all affect each other in subtle ways, and must be designed and understood together.
Specifically, there is an intricate tradeoff between the precision of subsumption, which determines the extent of pruning (and therefore the compactness of the representation), and the complexity of abstract domain operations such as weakening 
(e.g., for computing $a \join \alpha(\{\state\})$).
The definitions, algorithms, and data structures we present are carefully crafted to balance these considerations.
Our subsumption relation, which approximates entailment, is cheap to compute, eliminates enough redundancy to keep the representation of
abstract elements compact, and enables an efficient implementation of the weakening operation.

To evaluate our implementation of the abstract domain, we use it to implement a symbolic abstraction~\cite{symabs2004} procedure that computes the least fixpoint of the best abstract transformer of a transition system (i.e., %
the strongest inductive invariant for the transition system in the given language).
Our evaluation uses benchmarks from the literature, mostly from safety verification of distributed protocols.
While our fixpoint computation algorithm is not fully competitive with property-directed invariant inference approaches that exploit various sophisticated heuristics and optimizations,
it does demonstrate that fixpoint computation in our abstract domain is feasible,
which is quite surprising given the amount of quantified formulas the domain considers.
Our approach successfully
computes the least fixpoint
for transition systems that previously could only be analyzed using property-directed, heuristic techniques (which do not compute the least fixpoint, but an unpredictable heuristic fixpoint).
For example, we succeed at finding the strongest inductive invariant of Paxos as modeled in~\cite{paxos-made-epr} (which in our representation has 1,438 formulas with $\forall^*\exists^*\forall^*$ quantification, representing %
orders of magnitude more subsumed formulas).

In summary, this paper makes the following contributions:
\begin{compactenum}
  \item We develop 
  a compact representation of sets of formulas based on a novel syntactic \emph{subsumption relation}. We make a tradeoff here between the extent of pruning and efficiency, accepting some redundant formulas in exchange for practical algorithms. (\Cref{sec:representation})
  \item We show how to implement a key operation of \emph{weakening} a formula to be satisfied by a given state,
  and leverage it to compute the join of an abstract element and the abstraction of a state, when abstract elements are represented using our subsumption-based representation.
  (\Cref{sec:weaken})
  \item We present a data structure that provides an efficient implementation of operations used in the join computation described above.
  (\Cref{sec:datastructure})
  \item We evaluate the approach by applying it to compute the least fixpoint of the best abstract transformer for several distributed and concurrent protocols from the literature, demonstrating the promise of our approach.
  (\Cref{sec:eval})
\end{compactenum}

The rest of this paper is organized as follows:
\Cref{sec:background} introduces definitions and notation, \Cref{sec:representation,sec:weaken,sec:datastructure,sec:eval} present the main contributions outlined above,
\Cref{sec:related} discusses related work,
and \Cref{sec:conclusion} concludes.
The proofs of all theorems stated in the paper are given
\ifextended
in \Cref{app:proofs}. An extended running example appears in \Cref{app:running-example}.
\else
in~\cite{extended-version}.
\fi

\section{Background}
\label{sec:background}

\paragraph{First-order logic.}
For simplicity of the presentation, we present our approach for single-sorted first-order logic, although in practice we consider many-sorted logic.
The generalization of our methods to many-sorted logic is straightforward.

Given a first-order signature $\signature$ that consists of constant, function and relation symbols, the sets of terms and formulas are defined in the usual way: a \emph{term} $t$ is either a variable $x$, a constant $c$ or a function application $f(t_1,\ldots,t_n)$ on simpler terms; a \emph{formula} is either an equality between terms $t_1 = t_2$, a relation  application $r(t_1,\ldots,t_n)$ on terms, or the result of applying Boolean connectives or quantification. We also include $\bot$ as a formula (that is never satisfied).

Terms and formulas are interpreted over first-order structures and assignments to the (free) variables. Given a first-order signature $\signature$, a structure $\struct=(\univ,\interp)$ consists of a \emph{universe} $\univ$ and an \emph{interpretation} $\interp$ to the symbols in $\signature$. We denote by $\structs{\signature}$ the set of structures of $\signature$ whose universe is a finite set.%
\footnote{%
We restrict our attention to FOL fragments that have a finite-model property.}
When considering formulas with free variables $\vars$, and given some structure $\sigma=(\univ,\interp)$,
an \emph{assignment} $\asgn : \vars\to\univ$ maps each variable to an element of the structure's universe. We write $(\struct,\asgn)\models \phi$ to mean that a structure $\struct$ with an assignment $\asgn$ satisfies a formula $\phi$, and $\psi\models\phi$ to mean that a formula $\psi$ semantically entails $\phi$, i.e., $(\struct,\asgn)\models \psi$ whenever $(\struct,\asgn)\models \phi$.

\paragraph{Abstract interpretation.}
Abstract interpretation~\cite{cc77,cc79} is a framework for approximating the semantics of systems.
It assumes a concrete domain and an abstract domain, each given by a partially ordered set, $(\cdomain, \csqsubseteq)$ and  $(\adomain, \asqsubseteq)$, respectively.
These are related via a Galois connection consisting of a monotone abstraction function $\alpha : \cdomain \to \adomain$ and a monotone concretization function $\gamma : \adomain \to \cdomain$ satisfying
$\alpha(c)\asqsubseteq a \iff c\csqsubseteq \gamma(a)$ for all $a\in \adomain$ and $c\in\cdomain$.

In this work we consider logical abstract domains parameterized by a finite first-order language $\lang$ of closed formulas over signature $\signature$.
In this context,
concrete elements are sets of states from $\states = \structs{\signature}$,%
\footnote{Later we consider non-closed formulas and let $\states$ denote structures with assignments.}
i.e., $\cdomain=\powerset{\states}$,
ordered by $\csqsubseteq=\subseteq$ (set inclusion).
Abstract elements are sets of formulas from $\lang$,
i.e., $\adomain=\powerset{\lang}$, ordered by $\asqsubseteq=\supseteq$,
and the Galois connection is given by
$\alpha(S) = \{\phi\in\lang \mid \forall \state\in \states.\ \state\models\phi\}$ and
$\gamma(F) = \{\state\in \states \mid \forall \phi\in F.\ \state\models\phi\}$.
That is, abstraction in this domain consists of all $\lang$-formulas that hold on a given concrete set, and concretization consists of all states that satisfy a given set of formulas.
Note that sets of formulas are interpreted conjunctively in this context.

This logical abstract domain forms a join-semilattice (meaning every two elements have a least upper bound) with a least element.
The least element, denoted $\bot_\adomain$ (not to be confused with the formula $\bot$), is $\lang$, and join, denoted $\join$, corresponds to set intersection.
For example, $F \join \alpha(\{\state\}) = F \cap \{\phi \in \lang \mid \state\models\phi \} = \{\phi \in F \mid \state \models \phi\}$,
and can be understood as \emph{weakening} $F$ by eliminating from it all formulas that are not satisfied by $\state$.

\section{Subsumption-Based Representation of Sets of Formulas}
\label{sec:representation}

In this section we develop an efficient representation for elements in the abstract domain $\adomain = \powerset{\lang}$ induced by a finite first-order language $\lang$.
The abstract elements are sets of formulas, interpreted conjunctively, which may be extremely large (albeit finite).
Our idea is to reduce the size and complexity of such sets by avoiding redundancies that result from semantic equivalence and entailment. For example, when representing a set of formulas we would like to avoid storing both $\phi$ and $\psi$ when they are semantically equivalent ($\phi \equiv \psi$). Similarly, if $\phi \models \psi$
then instead of keeping both $\phi$ and $\psi$ we would like to keep only $\phi$.

In practice, it is not possible to remove all such redundancies based on semantic equivalence and entailment,
since, as we shall see in \Cref{sec:weaken}, performing operations over the reduced representation of abstract elements involves recovering certain subsumed formulas, and finding these in the case of entailment essentially requires checking all formulas in the language. This is clearly infeasible for complex languages such as the ones used in our benchmarks (see \Cref{tab:fixpoint}), and is exacerbated by the fact that merely checking entailment is expensive for formulas with quantifiers. Instead, our key idea is to remove redundancies based on a cheap-to-compute subsumption relation, which approximates semantic entailment, and enables efficient operations over abstract elements such as joining them with an abstraction of a concrete state.

We start the section with an inductive definition of a family of finite first-order languages that underlies all of our developments (\Cref{sec:inductive-constructors}).
We then introduce a syntactic \emph{subsumption} relation for first-order formulas (\Cref{sec:subsumption}), which we
leverage to develop an efficient
\emph{canonicalization} of formulas, effectively determining a single representative formula for each subsumption-equivalence class (\Cref{sec:canonicalization}). We then use antichains of canonical formulas, i.e., sets of canonical formulas where no formula is subsumed by another, to represent sets of formulas (\Cref{sec:representing-sets}). \Cref{sec:weaken,sec:datastructure} develop ways to effectively manipulate this representation in order to accommodate important operations for abstract interpretation algorithms, such as weakening an abstraction to include a given concrete state.

\subsection{Bounded First-Order Languages}
\label{sec:inductive-constructors}

At core of our approach is an inductively-defined family of first-order languages, termed \emph{bounded first-order languages}.
These languages are all finite and bound various syntactic measures of formulas (e.g., number of quantifiers, size of the Boolean structure), which, in turn, determine the precision of the abstract domain.
The inductive definition of bounded languages facilitates efficient recursive implementations of our developments.

We fix a signature $\signature$ and a variable set $\vars$.
\Cref{def:lang} provides the inductive definition of the family of bounded first-order languages (over $\signature$ and $V$),
 where each language $\lang$ is also equipped with a bottom element $\bot_\lang$ (equivalent to false).
We use $\mathfrak{S}_X$ to denote the set of permutations over a set of variables $X$,
and use $\phi \pi$ to denote
the formula obtained by substituting free variables in a formula $\phi$ according to $\pi \in \mathfrak{S}_X$.
A set of formulas $F$ is $\mathfrak{S}_X$-closed if $\phi \pi \in F$ for every $\phi \in F$, $\pi \in \mathfrak{S}_X$.
All bounded first-order languages will be $\mathfrak{S}_V$-closed; this will be important for canonicalization.
We use $\tupphi = \seq{\phi_1, \ldots, \phi_n}$ to denote a sequence of formulas,
$\phi_{-i}$ to denote the formula $\phi_{n-i+1}$ in the sequence,
$\card{\tupphi}$ for the length of
$\tupphi$, and
$\indices{\tupphi}$ for its set of indices $\{1,\dots,|\tupphi|\}$.
We  use $\lang^*$ for the set of all (finite) sequences of formulas from $\lang$,
and $\emptytup$ for the empty sequence ($\card{\emptytup}=0$).

\begin{definition}[Bounded First-Order Languages]
\label{def:lang}
A \emph{bounded first-order language} is one of the following, where
$X \subseteq \vars$ denotes a finite set of variables, and
$\lang$, $\lang_1$ and $\lang_2$ denote bounded first-order languages:
\begin{footnotesize}
\begin{align*}
\langa &= \aset \cup \{\bot\}
\text{ with } \bot_{\langa}=\bot
\text{, where } \aset \text{ is any finite $\mathfrak{S}_\vars$-closed set of formulas}
\\
\orAB{\lang_1}{\lang_2} &= \{\phi_1\vee\phi_2 \mid \phi_1\in\lang_1, \phi_2\in\lang_2\}
\text{ with } \bot_{\orAB{\lang_1}{\lang_2}} = \bot_{\lang_1} \vee \bot_{\lang_2}
\\
\andAB{\lang_1}{\lang_2} &= \{\phi_1\wedge\phi_2 \mid \phi_1\in\lang_1, \phi_2\in\lang_2\}
\text{ with } \bot_{\andAB{\lang_1}{\lang_2}} = \bot_{\lang_1} \wedge \bot_{\lang_2}
\\
\orA{k}{\lang} &= \{\bigvee \tupphi \mid \tupphi\in\lang^* \text{ and } \card{\tupphi} \leq k\}
\text{ with } \bot_{\orA{k}{\lang}} = \bigvee \emptytup
\text{, where } k \in \mathbb{N}
\\
\andA{\omega}{\lang} &= \{\bigwedge \tupphi \mid \emptytup \neq \tupphi \in \lang^*  \}
\text{ with } \bot_{\andA{\omega}{\lang}} = \bigwedge
\seq{\bot_\lang}
\\
\existsA{X}{\lang} &=\{\exists X.\phi \mid \phi\in\lang\}
\text{ with } \bot_{\existsA{X}{\lang}} = \exists X. \bot_\lang
\\
\forallA{X}{\lang} &= \{\forall X.\phi \mid \phi\in\lang\}
\text{ with } \bot_{\forallA{X}{\lang}} = \forall X. \bot_\lang
\\
\existsforallA{X}{\lang} &= \left\{\quant X.\phi \mid \phi\in\lang, \quant \in \left\{\exists, \forall\right\}\right\}
\text{ with } \bot_{\existsforallA{X}{\lang}} = \forall X. \bot_\lang
\end{align*}
\end{footnotesize}
\end{definition}

The base case is any finite set of formulas (over $\signature$ and $V$) that is closed under variable permutations, augmented by $\bot$ (denoting false). Typical examples include the set of all literals over $\signature$ and $V$ with a bounded depth of function applications.
We introduce binary language constructors for disjunction and conjunction, each operating on two possibly different languages.
We also introduce constructors for homogeneous disjunction of at most $k$ disjuncts,
as well as unbounded non-empty conjunction, over any single language.
Finally, we introduce constructors for quantification ($\exists$ or $\forall$) over a finite set of variables and a language,
as well as a constructor that includes both quantifiers for languages where both options are desired.
Note that for the construction of a logical abstract domain, we are interested in languages where all formulas are closed (have no free variables), but the inductive definition includes
languages with free variables.

The semantics of formulas in each language is defined w.r.t. states $\states$ that consist of first-order structures and assignments to the free variables, following the standard first-order semantics, extended to conjunctions and disjunctions of finite sequences in the natural way, where $\bigvee \emptytup \equiv \bot$.
(We do not allow $\bigwedge \emptytup$, which would have been equivalent to ``true'',  since it is not useful for our developments.)

Observe that for a fixed language $\lang$, the formulas $\phi_1\vee\phi_2 \in \orAB{\lang}{\lang}$ and $\bigvee \seq{\phi_1,\phi_2} \in \orA{2}{\lang}$ are syntactically different but semantically equivalent (and similarly for conjunctions).
Nonetheless, we introduce homogeneous disjunction and conjunction since they admit a more precise subsumption relation, yielding a more efficient representation of sets of formulas.
Also note that we consider bounded disjunction but unbounded conjunction; \Cref{sec:design} explains this choice.

\begin{example} \label{ex:bounded-lang}
$\lang = \forallA{\{x,y\}}{\orA{2}{\lang_A}}$ with $A = \{ p(x),\neg p(x), p(y),\neg p(y) \}$ is a bounded first-order language over signature $\signature$ that has one unary predicate $p$ and variables $V=\{x,y\}$.
Formulas in this language are universally quantified homogeneous disjunctions of at most two literals. For instance, $\lang$ includes $\forall\{x,y\}.\bigvee\epsilon$, which is also $\bot_\lang$, as well as $\forall\{x,y\}.\bigvee\seq{p(x)}$,  $\forall\{x,y\}.\bigvee\seq{p(x), \neg p(y)}$, etc.
\end{example}

\subsection{Syntactic Subsumption}
\label{sec:subsumption}

Next, we define a \emph{subsumption} relation for each bounded first-order language. The subsumption relation serves as an easy-to-compute under-approximation for  entailment between formulas from the same language.
We use $\sqsubseteq_\lang$ to denote the subsumption relation for language $\lang$, or simply $\sqsubseteq$ when $\lang$ is clear from context. When $\phi \sqsubseteq \psi$ we say $\phi$ \emph{subsumes} $\psi$, and then we will also have $\phi \models \psi$.

\begin{definition}[Subsumption]
  \label{def:subsumption}
We define $\sqsubseteq_\lang$ inductively, following the definition of bounded first-order languages, as follows, where $\veewedge \in \{\vee, \wedge\}$,  $k\in\mathbb{N}$,
$\quant, \quant' \in \{\exists, \forall\}$, $X$ is a finite set of variables,
and $\lang$, $\lang_1$ and $\lang_2$ are bounded first-order languages:
\begin{footnotesize}
\begin{align*}
\phi \sqsubseteq_{\langa} \psi &\text{ iff } \phi = \bot \text{ or } \phi = \psi
\\
\phi_1\veewedge\phi_2 \sqsubseteq_{\orandAB{\lang_1}{\lang_2}} \psi_1\veewedge\psi_2
&\text{ iff }
\phi_1\sqsubseteq_{\lang_1}\psi_1 \text{ and } \phi_2\sqsubseteq_{\lang_2}\psi_2
\quad\text{(pointwise extension)}
\\
\bigvee \tupphi \sqsubseteq_{\orA{k}{\lang}} \bigvee \tuppsi
&\text{ iff }
\exists m \colon \indices{\tupphi} \to \indices{\tuppsi}. \,
\forall i \in \indices{\tupphi}. \, \phi_i \sqsubseteq \psi_{m(i)}
\text{ and } m \text{ is injective}
\\
\bigwedge \tupphi \sqsubseteq_{\andA{\omega}{\lang}} \bigwedge \tuppsi &\text{ iff }
\exists m \colon \indices{\tuppsi} \to \indices{\tupphi}. \,
\forall i \in \indices{\tuppsi}. \, \phi_{m(i)} \sqsubseteq \psi_i
\\
(\quant X.\phi) \sqsubseteq_{\quantA{X}{\lang}} (\quant X.\psi)
&\text{ iff }
\exists \pi \in \mathfrak{S}_X. \, \phi\sqsubseteq_\lang \psi \pi
\\
(\quant X.\phi) \sqsubseteq_{\existsforallA{X}{\lang}} (\quant' X.\psi)
&\text{ iff }
\exists \pi \in \mathfrak{S}_X. \, \phi\sqsubseteq_\lang \psi \pi
\text{, and }
\quant = \forall \text{ or } \quant' = \exists
\end{align*}
\end{footnotesize}
\end{definition}

The subsumption relation of a bounded first-order language $\lang$
is composed, hierarchically, from the subsumption relations of the bounded first-order languages that $\lang$ is composed from.
For example,
the languages participating in the composition of $\lang = \forallA{\{x,y\}}{\orA{2}{\lang_A}}$ defined in \Cref{ex:bounded-lang} are $\lang_A$, $\orA{2}{\lang_A}$, and $\forallA{\{x,y\}}{\orA{2}{\lang_A}}$, and each is equipped with its own subsumption relation.

In the base case, formulas in ${\langa}$ are only subsumed by themselves or by $\bot$.
For example, considering \Cref{ex:bounded-lang}, $p(x)\not\sqsubseteq_{\lang_A} p(y)$.
Subsumption is lifted to languages obtained by binary conjunctions and disjunctions in a pointwise manner.
For the languages obtained by homogeneous constructors,
a mapping over indices determines which element of one sequence subsumes which element of the other.
To approximate entailment, the mapping in the disjunctive case maps each element of $\bigvee \tupphi$ to one in $\bigvee \tuppsi$ that it subsumes, and in the conjunctive case maps each element of $\bigwedge \tuppsi$ to one in $\bigwedge \tupphi$ that subsumes it.
As a result, subsumption is more precise in the homogeneous case than in the binary one.
For example, considering $A$ from \Cref{ex:bounded-lang},
$p(x) \lor p(y) \not\sqsubseteq_{\orAB{\lang_A}{\lang_A}} p(y) \lor p(x)$, even though the formulas are semantically equivalent.
On the other hand, $\bigvee \seq{p(x), p(y)} \sqsubseteq_{\orA{2}{\lang_A}} \bigvee \seq{p(y), p(x)}$%

In the case of quantifiers, subsumption is lifted from %
the language of the body while considering permutations over the quantified variables.
For example, in \Cref{ex:bounded-lang}, $\forall\{x,y\}.\bigvee\seq{p(x)} \sqsubseteq_{\lang} \forall\{x,y\}.\bigvee\seq{p(y)}$ due to variable permutations, even though $\bigvee\seq{p(x)} \not\sqsubseteq_{\orA{2}{\lang_A}} \bigvee\seq{p(y)}$.
When both quantifiers are considered, a universal quantifier can subsume an existential one.

The injectivity requirement for $\sqsubseteq_{\orA{k}{\lang}}$ can be dropped without damaging any of the definitions or theorems in this section, but it enables a simpler definition of the weakening operator in \Cref{sec:weaken} (as discussed further in \Cref{sec:design}).

The following theorem establishes the properties of $\sqsubseteq_\lang$.

\begin{theorem}[Properties of $\sqsubseteq_\lang$]
	\label{thm:subsumption}
    For any bounded first-order language $\lang$, $\sqsubseteq_\lang$ is a preorder (i.e., reflexive and transitive) such that for any $\phi,\psi\in\lang$, if $\phi \sqsubseteq \psi$ then $\phi \models \psi$. Moreover, $\bot_\lang \sqsubseteq_\lang \phi$ for any $\phi \in \lang$.
\end{theorem}

As with entailment, where two distinct formulas can entail each other (i.e., be semantically equivalent), there can be distinct formulas $\phi,\psi \in \lang$ with $\phi \sqsubseteq_\lang \psi$ and $\psi \sqsubseteq_\lang \phi$ (since  $\sqsubseteq_\lang$ is not always a partial order, i.e., not  antisymmetric).
We call such formulas \emph{subsumption-equivalent}, and denote this by $\phi \equiv_{\sqsubseteq_\lang} \psi$. ($\equiv_{\sqsubseteq_\lang}$ is clearly an equivalence relation.)
The existence of subsumption-equivalent formulas is a positive sign, indicating that our subsumption relation manages to capture nontrivial semantic equivalences.  
This is thanks to
the definition of subsumption for homogeneous disjunction and conjunction,
as well as for quantification.
For example, $\bigvee \seq{\phi,\psi} \equiv_\sqsubseteq \bigvee \seq{\psi,\phi}$ (and similarly for conjunction), and if $\phi \sqsubseteq \psi$ then
$\bigwedge \seq{\phi, \psi} \equiv_\sqsubseteq \bigwedge \seq{\phi}$.
For quantifiers, $\quant X. \phi \equiv_\sqsubseteq \quant X. \phi \pi$ for any $\pi \in \mathfrak{S}_X$ and $\quant \in \{\exists, \forall\}$.
(In contrast, $\sqsubseteq_{\langa}$ is always antisymmetric, and the definitions of $\orAB{\lang_1}{\lang_2}$ and $\andAB{\lang_1}{\lang_2}$ preserve antisymmetry.)

\subsection{Canonicalization}
\label{sec:canonicalization}

As a first step towards an efficient representation of sets of formulas,
we use a canonicalization of formulas w.r.t. $\equiv_\sqsubseteq$,
which allows us to only store canonical formulas as unique representatives of their (subsumption-) equivalence class.
In general, a \emph{canonicalization} w.r.t. an equivalence relation $\equiv$ over a set $S$ is a function $c\colon S \to S$
such that
$\forall x \in S.\, c(x) \equiv x$ (representativeness) and
$\forall x,y \in S.\, x \equiv y \iff c(x) = c(y)$ (decisiveness).
We say that $x$ is \emph{canonical} if $c(x) = x$. When the equivalence relation is derived from a preorder (as $\equiv_\sqsubseteq$ is derived from $\sqsubseteq$) then the preorder is a partial order over the set of canonical elements.
For our case, that means that $\sqsubseteq_\lang$ is a partial order over the set of canonical $\lang$-formulas.

It is useful, both for the algorithms developed in the sequel and for the definition of canonicalization for $\quantA{X}{\lang}$ ($Q \in \{\bm\exists, \bm\forall, \existsforall\}$), to define a total order $\leq_\lang$ over canonical $\lang$-formulas that extends $\sqsubseteq_\lang$. We thus define the canonicalization function $c_\lang$ and the total order $\leq_\lang$ over canonical $\lang$-formulas by mutual induction.
For a set of canonical $\lang$-formulas $F$, we use $\min_{\sqsubseteq_\lang} F$ to denote the set of formulas in $F$ not subsumed by others,
i.e., $\min_{\sqsubseteq_\lang} F = \{\phi \in F \mid \forall \psi \in F \setminus \{\phi\}. \, \psi \not\sqsubseteq \phi  \}$,
and use $\min_{\leq_\lang} F$ to denote the minimal element of a non-empty set $F$ w.r.t. the total order $\leq_\lang$.
Finally, we use $\ordered{\leq}{\tupphi}$ for the sequence obtained by sorting $\tupphi$ according to $\leq$ in ascending order,
and similarly $\ordered{\leq}{F}$ for the sequence obtained by sorting the elements of a set $F$.

\begin{definition}[Canonicalization]
\label{def:canonicalization}
  For every bounded first-order language $\lang$, we define the \emph{canonicalization function} $c_\lang\colon \lang \to \lang$ and a total order $\leq_\lang$ over canonical $\lang$-formulas
  by mutual induction
  (where $\veewedge \in \{\vee, \wedge\}$ and $\quant \in \{\exists, \forall\}$):
\begin{footnotesize}
\begin{align*}
c_{\langa}(\phi) &= \phi
\\
c_\orandAB{\lang_1}{\lang_2}(\phi_1\veewedge\phi_2) &= c_{\lang_1}(\phi_1) \veewedge c_{\lang_2}(\phi_2)
\text{ (pointwise)}
\\
c_\orA{k}{\lang}(\bigvee \tupphi) &= 
 \bigvee \ordered{\leq_\lang}{ c_\lang(\phi_1),\ldots,c_\lang(\phi_{\card{\tupphi}})}
\\
c_\andA{\omega}{\lang}(\bigwedge \tupphi) &= \bigwedge \ordered{\leq_\lang}{\min{}_{\sqsubseteq_\lang} \{ c_\lang(\phi_1),\ldots,c_\lang(\phi_{\card{\tupphi}})\}}
\\
c_\quantA{X}{\lang}(\quant X.\phi) &= \quant X. \min{}_{\leq_\lang} \left\{c_\lang(\phi \pi) \bigm| \pi \in \mathfrak{S}_X\right\}
\\
c_\existsforallA{X}{\lang}(\quant X.\phi) &= c_\quantA{X}{\lang}(\quant X.\phi)
\end{align*}
\end{footnotesize}
and %
\begin{footnotesize}
\begin{align*}
\leq_{\langa} &\text{ is an arbitrary total order extending } \sqsubseteq_{\langa}
\\
\phi_1 \veewedge \phi_2 \leq_{\orandAB{\lang_1}{\lang_2}} \psi_1\veewedge\psi_2
&\iff
\phi_1 <_{\lang_1} \psi_1 \text{, or } \phi_1=\psi_1 \text{ and } \phi_2 \leq_{\lang_2} \psi_2
\\
\begin{split}
\bigvee \tupphi \leq_{\orA{k}{\lang}} \bigvee \tuppsi
&\iff
\tupphi \text{ is a suffix of }\tuppsi, \\
&\hspace{3.5em}\text{or } \exists i \in
\indices{\tupphi}\cap\indices{\tuppsi}.
\  \phi_{-i} <_\lang \psi_{-i} \land \forall j < i.\ \phi_{-j} = \psi_{-j} 
\end{split}
\\
\begin{split}
\bigwedge \tupphi \leq_{\andA{\omega}{\lang}} \bigwedge \tuppsi
&\iff
\tuppsi \text{ is a prefix of }\tupphi, \\
&\hspace{3.5em}\text{or } \exists i \in
\indices{\tupphi}\cap\indices{\tuppsi}.
\  \phi_i <_\lang \psi_i \land \forall j < i.\ \phi_j = \psi_j 
\end{split}
\\
\quant X.\phi \leq_{\quantA{X}{\lang}} \quant X.\psi
&\iff \phi\leq_\lang\psi
\\
\quant X.\phi \leq_{\existsforallA{X}{\lang}} \quant' X.\psi
&\iff Q=Q' \text{ and } \phi\leq_\lang\psi \text{, or } Q=\forall \text{ and } Q'=\exists
\end{align*}
\end{footnotesize}

where $\phi <_\lang \psi$ is shorthand for ``$\phi \leq_\lang \psi$ and $\phi \neq \psi$''.
\end{definition}

Our inductive definition of canonicalization in \Cref{def:canonicalization} recognizes the only possible sources of nontrivial subsumption-equivalence in our construction: non-canonicity of subformulas, ordering of sequences, internal subsumption in $\andA{\omega}{\cdot}$-sequences, and permuting of quantified variables. To address these, we canonicalize all subformulas, order their sequences w.r.t $\leq_\lang$ in $\orA{k}{\lang}$ and $\andA{\omega}{\lang}$,
minimize $\andA{\omega}{\lang}$-sequences w.r.t $\sqsubseteq_\lang$, and in $\quantA{X}{\lang}$, $Q\in\{\bm\exists,\bm\forall,\existsforall\}$, choose the permutation yielding the $\leq_\lang$-least (canonical) body. For the total order in the cases of Boolean connectives, we use lexicographic-like orderings carefully designed to extend their associated subsumption relations (e.g., homogeneous disjunction uses a right-to-left lexicographic ordering).
For quantification, the total order is directly lifted from the total order for canonical bodies.

As an example, consider $\lang = \forallA{\{x,y\}}{\orA{2}{\lang_A}}$ from \Cref{ex:bounded-lang}. To obtain a canonicalization for $\lang$, we provide an arbitrary total order $\leq_{\lang_A}$, say
$p(x) <_{\lang_A} \neg p(x) <_{\lang_A} p(y) <_{\lang_A} \neg p(y)$ (recall that $\bot\in\lang_A$ is least). This uniquely determines the total order and canonicalization of $\lang$ and all of its sub-languages.
For example, canonicalization of both
$\forall\{x,y\}.\bigvee\seq{p(x)}$ and
$\forall\{x,y\}.\bigvee\seq{p(y)}$, which are $\sqsubseteq_\lang$-equivalent, is $\forall\{x,y\}.\bigvee\seq{p(x)}$.
This is because $p(x)<_{\lang_A}p(y)$, and thus
$c_\orA{2}{\lang_A}\left(\bigvee\seq{p(x)}\right) = \bigvee \seq{p(x)} <_\orA{2}{\lang_A} \bigvee\seq{p(y)} = c_\orA{2}{\lang_A}\left(\bigvee\seq{p(y)}\right)$.
Note that $\bigvee\seq{p(x)}$ and $\bigvee\seq{p(y)}$ are both canonical, but adding quantifiers merges the two formulas into the same subsumption-equivalence class, necessarily making the quantified version of one of them non-canonical.
Similarly, the $\sqsubseteq_{\orA{2}{\lang_A}}$-equivalent formulas  $\bigvee\seq{p(x),p(y)}$ and $\bigvee\seq{p(y),p(x)}$ are both canonicalized into $\bigvee\seq{p(x),p(y)}$ (by sorting the sequences of literals according to $\leq_{\lang_A}$).

The properties of $c_\lang$ and $\leq_\lang$ defined above
are established by the following theorem,
which ensures that \Cref{def:canonicalization} is well-defined (e.g., that whenever $\min_{\leq_\lang}$ is used, $\leq_\lang$ is a total order).

\begin{theorem}
    \label{thm:canonicalization}
    For any bounded language $\lang$,
    $c_\lang$ is a canonicalization w.r.t. $\equiv_{\sqsubseteq_\lang}$,
    that is, it is
    representative ($c_\lang(\phi) \equiv_{\sqsubseteq_\lang} \phi$)
    and decisive ($\phi \equiv_{\sqsubseteq_\lang} \psi \iff c_\lang(\phi) = c_\lang(\psi)$);
    $\sqsubseteq_\lang$ is a partial order over canonical $\lang$-formulas; and
    $\leq_\lang$ is a total order over canonical $\lang$-formulas that extends $\sqsubseteq_\lang$.
\end{theorem}

\begin{corollary}
    For any $\phi,\psi \in \lang$, if $\phi\sqsubseteq_\lang\psi$ then $c_\lang(\phi) \leq_\lang c_\lang(\psi)$.
\end{corollary}

Henceforth, we use $\clang{\lang}$ to denote the set of canonical $\lang$-formulas. %

\subsection{Representing Sets of Formulas}
\label{sec:representing-sets}

We utilize the subsumption relation and canonicalization to efficiently represent sets of formulas which are interpreted conjunctively as antichains of canonical formulas, where an \emph{antichain} is a set of formulas incomparable by subsumption.

\begin{definition}[Set Representation]
Given a set of formulas $F \subseteq \lang$, we define its \emph{representation} as the set $R_F = \min{}_{\sqsubseteq_{\lang}} \{c(\phi) \mid \phi \in F\}$.
\end{definition}

The representation combines two forms of redundancy elimination:
the use of canonical formulas eliminates redundancies due to subsumption-equivalence,
and
the use of $\sqsubseteq_{\lang}$-minimal elements reduces the size of the set by ignoring subsumed formulas.
Observe that the more permissive the subsumption relation is, the smaller the set representations are,
because more formulas will belong to the same equivalence class and more formulas will be dropped by $\min_{\sqsubseteq_\lang}$.

This representation preserves the semantics of a set of formulas (interpreted conjunctively).
For sets that are upward-closed w.r.t. subsumption (e.g., $\alpha(S)$ for some set of states $S$), the representation is lossless as a set can be recovered by taking the upward closure of its representation.
For a set $F \subseteq \lang$, we use $\upclose{F}$ to denote its \emph{upward closure} (w.r.t. $\sqsubseteq_\lang$),
given by $\upclose{F} = \{\phi \in \lang \mid \exists \psi \in F. \, \psi \sqsubseteq_\lang \phi\}$.

\begin{theorem}[Antichain Representation]
\label{thm:representation}
For $F \subseteq \lang$ and $R_F = \min{}_{\sqsubseteq_{\lang}} \{c(\phi) \mid \phi \in F\}$ its representation,
$\bigwedge R_F \equiv \bigwedge F$ and
$\upclose{R_F} = \upclose{F}$.
\end{theorem}

\begin{corollary}
  \label{cor:antichain}
  If $F \subseteq \lang$ is upward closed w.r.t.\ $\sqsubseteq_\lang$ then $F = \upclose{R_F}$.
\end{corollary}

In particular, \Cref{cor:antichain} applies to any set that is closed under entailment.

\section{The Weaken Operator}
\label{sec:weaken}

This section develops an algorithm that computes a \emph{weaken} operator, which takes a representation of an upward-closed set $F\subseteq\lang$ and a state $s$ and computes a representation of $F \cap \alpha(\{\state\}) = \{\phi \in F \mid \state \models \phi\}$.
When $F$ is viewed as an abstract element, this operation corresponds to computing $F \join \alpha(\{\state\})$.
While it is not a general abstract join operator,
joining an abstract element with the abstraction of a single concrete state
is a powerful building block that can be used, for example, to compute the abstraction of a set of states or even
the least fixpoint of the best abstract transformer (\emph{\'a la} symbolic abstraction~\cite{symabs2004}).

In an explicit representation of $F$, computing $F \join \alpha(\{\state\})$ would amount to removing from $F$ all the formulas that are not satisfied by $\state$. However, in the subsumption-based representation $R_F$, simply removing said formulas is not enough. Instead, we must \emph{weaken} them, i.e., replace them by formulas they subsume that are satisfied by $\state$.
To this end, \Cref{sec:weaken-formula} develops an appropriate weakening operator for a single formula,
and \Cref{sec:weaken-set} then lifts it to antichains used as representations.

\subsection{Weakening a Single Canonical Formula}
\label{sec:weaken-formula}

Given a canonical formula $\phi$ and a state $\state$ such that $\state \not\models \phi$, the weaken operator computes the set of minimal canonical formulas that are subsumed by $\phi$ and satisfied by  $\state$,
which can be understood as a representation of $\upclose{\{\phi\}}\cap\alpha(\{\state\})$.

\begin{definition}[The Weaken Operator]
    \label{def:weaken-operator-lang}
    The \emph{weaken operator} of $\lang$ is the function $\weakenop_\lang \colon \clang{\lang} \times \states \to \powerset{\clang{\lang}}$
    defined as follows:
    \begin{equation*}
       \weaken{\lang}{\phi}{\state}= \min{}_{\sqsubseteq_\lang} \left\{c_\lang(\psi) \mid \psi\in\lang,  \phi\sqsubseteq\psi \text{, and } \state\models\psi\right\}.
    \end{equation*}
\end{definition}

Note that $\weaken{\lang}{\phi}{\state}$ returns a set of formulas, since there may be different incomparable ways to weaken $\phi$ such that it is satisfied by $\state$.

While \Cref{def:weaken-operator-lang} does not suggest a way to compute $\weaken{\lang}{\phi}{\state}$,
the following theorem provides a recursive implementation of $\weaken{\lang}{\phi}{\state}$ that follows
the inductive structure of bounded languages.
For the quantification cases, we weaken according to all assignments of variables in $X \subseteq V$.
Recall that a state can be unpacked as $\state = ((\univ,\interp),\asgn)$ where $(\univ,\interp)$ is a first-order structure (universe and interpretation) and $\asgn$ is an assignment to variables (into $\univ$).
For assignments $\asgn$ and $\asgnb$, we use $\stateupdate{\asgn}{\asgnb}$ to denote the assignment obtained from $\asgn$ by updating (possibly extending) it according to $\asgnb$.

\begin{theorem}[Implementation of Weaken]
\label{thm:weaken-implementation}
Let $\phi \in \lang$ be a canonical formula in a bounded first-order language $\lang$ and $\state \in \states$ a state. If $\state \models \phi$ then $\weaken{\lang}{\phi}{\state} = \{ \phi \}$.
If $\state \not\models \phi$, then $\weaken{\lang}{\phi}{\state}$
is given by: %

\begin{footnotesize}
\begin{align*}
\weaken{\langa}{\phi}{\state} &=
\begin{cases}
    \{\psi \in \aset \mid s \models \psi\}, & \text{ if } \phi = \bot\\
    \emptyset, & \text{ if } \phi \neq \bot
\end{cases}
\\
\begin{split}
\weaken{\orAB{\lang_1}{\lang_2}}{\phi_1 \vee \phi_2}{\state} &=
\{\psi \vee \phi_2 \mid \psi \in \weaken{\lang_1}{\phi_1}{\state}\} \,
\cup \{\phi_1 \vee \psi \mid \psi \in \weaken{\lang_2}{\phi_2}{\state}\}
\end{split}
\\
\weaken{\andAB{\lang_1}{\lang_2}}{\phi_1 \wedge \phi_2}{\state} &=
\{\psi_1 \wedge \psi_2 \mid
\psi_1 \in \weaken{\lang_1}{\phi_1}{\state}, %
\psi_2 \in \weaken{\lang_2}{\phi_2}{\state}\}
\\
\begin{split}
\weaken{\orA{k}{\lang}}{\bigvee \tupphi}{\state} &= \min{}_{\sqsubseteq_\orA{k}{\lang}} \left( W_{\card{\tupphi}} \cup W_{\card{\tupphi}+1} \right) \text{ where } \\
&\hspace{-9em}
W_{\card{\tupphi}} =
\left\{\bigvee \ordered{\leq_\lang}{\phi_1,\ldots,\phi_{i-1},\psi,\phi_{i+1},\ldots,\phi_{\card{\tupphi}}} \mid
i\in\indices{\tupphi}, \psi \in \weaken{\lang}{\phi_i}{s}\right\} \text{ and } \\
&\hspace{-9em}
W_{\card{\tupphi}+1} =
\left\{\bigvee\ordered{\leq_\lang}{\phi_1,\ldots,\phi_{\card{\tupphi}},\psi} \mid \psi \in \weaken{\lang}{\bot_\lang}{s} \text{ and }
\card{\tupphi} < k
\right\}
\end{split}
\\
\weaken{\andA{\omega}{\lang}}{\bigwedge \tupphi}{\state} &=
\left\{ \bigwedge \ordered{\leq_\lang}{\min{}_{\sqsubseteq_\lang} \weaken{\lang}{\phi_1}{\state} \cup \cdots \cup \weaken{\lang}{\phi_\card{\tupphi}}{\state}} \right\}
\\
\begin{split}
\weaken{\existsA{X}{\lang}}{\exists X. \phi}{((\univ,\interp),\asgn)} &=
\min{}_{\sqsubseteq_\existsA{X}{\lang}}
\{ c(\exists X. \psi) \mid \asgnb\colon X \to \univ, \psi \in \weaken{\lang}{\phi}{((\univ,\interp),\stateupdate{\asgn}{\asgnb})} \}
\end{split}
\\
\begin{split}
\weaken{\forallA{X}{\lang}}{\forall X. \phi}{((\univ,\interp),\asgn)} &= 
\min{}_{\sqsubseteq_\forallA{X}{\lang}}
\{ c(\forall X. \psi) \mid \psi \in \Omega_\phi\left(\{((\univ,\interp),\stateupdate{\asgn}{\asgnb}) \mid \asgnb\colon X \to \univ\}\right) \} \\
&\hspace{-8em}\text{where }
\Omega_{\phi_0}\left(\{\state_1,\ldots,\state_n\}\right) =
\{\phi_n \mid
\phi_1 \in \weaken{\lang}{\phi_0}{\state_1},
\ldots,
\phi_n \in \weaken{\lang}{\phi_{n-1}}{\state_n}
\}
\end{split}
\\
\begin{split}
\weaken{\existsforallA{X}{\lang}}{\exists X. \phi}{\state} &= \weaken{\existsA{X}{\lang}}{\exists X. \phi}{\state} \\
\weaken{\existsforallA{X}{\lang}}{\forall X. \phi}{\state} &= \min{}_{\sqsubseteq_\existsforallA{X}{\lang}}\left(\weaken{\existsA{X}{\lang}}{\exists X. \phi}{\state} \cup \weaken{\forallA{X}{\lang}}{\forall X. \phi}{\state}\right)
\end{split}
\end{align*}
\end{footnotesize}
\end{theorem}

When $\state \models \phi$, no weakening of $\phi$ is needed for $\state$ to satisfy it.
In the case of $\langa$, only $\bot$ can be weakened to make $\state$ satisfy it, yielding the set of formulas from $\aset$ that are satisfied by $\state$. (For $\langa$, weakening anything except $\bot$ that is not satisfied by $\state$ yields the empty set.)
In the case of disjunction, it suffices for one of the disjuncts to be satisfied by $\state$.
Therefore, weakening
is done by %
(i)~weakening exactly one of the existing disjuncts, which applies to both $\orAB{\lang_1}{\lang_2}$ and $\orA{k}{\lang}$;
or by (ii)~adding a disjunct that weakens $\bot_{\lang}$, which applies only to $\bigvee \tupphi \in \orA{k}{\lang}$ when $\card{\tupphi} < k$.
In the case of homogeneous disjunction,  each resulting disjunction needs to be sorted to restore canonicity;
moreover, some of the resulting disjunctions may be subsumed by others,
so $\min{}_{\sqsubseteq_{\orA{k}{\lang}}}$ is applied to the set of weakened disjunctions.
In the case of conjunction, all conjuncts need to be weakened to be satisfied by $\state$.
In the binary case, this leads to all pairs that combine weakened conjuncts.
But in the homogeneous case a single conjunction can accumulate all weakened conjuncts, so weakening always yields a singleton set;
filtering the weakened conjuncts using $\min{}_{\sqsubseteq_\lang}$ is required to ensure canonicity, as one weakened conjunct may subsume another. %
To satisfy an existentially quantified formula, it suffices for the body to be satisfied by a single assignment.
Therefore, each possible assignment $\asgnb$ contributes to the result of weakening.  %
In contrast, for a universally quantified formula the body must be satisfied by all assignments.
Therefore, the body of the formula is iteratively weakened by \emph{all} assignments.
In both cases, formulas are re-canonicalized and non-minimal elements are removed.
The case of $\existsforallA{X}{\lang}$ combines the two quantified cases.

\begin{example}
Consider applying the weaken operator of $\lang = \forallA{\{x,y\}}{\orA{2}{\lang_A}}$ from \Cref{ex:bounded-lang} to the bottom element $\bot_\lang = \forall \{x,y\}.\bigvee \epsilon$, with the state $s=((\univ,\interp),\mu)$ where $\univ=\{a,b\}$, $p^\interp=\{a,b\}$, and $\mu$ is an empty assignment. 
To weaken the universally quantified formula, we first iteratively weaken its body, 
$\phi_0=\bigvee\epsilon$, with the states  $s_1,\ldots,s_4$, each of which  extends $s$ with one of the 4 possible assignments to $x,y$. 
Since all of these states satisfy $p(x)$ and $p(y)$, the first weakening (with $s_1$) results in $\{\bigvee \seq{p(x)}, \bigvee \seq{p(y)}\}$, and no formula is weakened further in later iterations (since both of them are already satisfied by $s_2,s_3,s_4$).  As we have seen in \Cref{sec:canonicalization}, both formulas are canonical; however, they become subsumption-equivalent when the quantifier prefix is added, demonstrating the need 
for additional canonicalization in the computation of weaken for $\forallA{X}{\cdot}$. The result is the antichain of canonical formulas $\{\forall\{x,y\}.\bigvee\seq{p(x)}\}$.
Note that the weakened formula $\bot_\lang$ has 21 formulas in its $\sqsubseteq_{\lang}$-upward closure,
and its weakening has 14 formulas
\ifextended
(see \Cref{app:example-weaken}
\else
(see~\cite{extended-version}
\fi
for the lists of formulas); yet throughout the weakening process we only dealt with at most two formulas.
\end{example}

\subsection{Weakening Sets of Formulas}
\label{sec:weaken-set}

We
lift the weaken operator
to sets of canonical formulas.
For a set $R \subseteq \clang{\lang}$, we define $\weaken{\lang}{R}{\state} = \min{}_\sqsubseteq \bigcup_{\phi \in R} \weaken{\lang}{\phi}{\state}$, 
\nolinebreak 
motivated
\nolinebreak 
by
\nolinebreak
the
\nolinebreak
following
\nolinebreak 
theorem.

\begin{theorem}[From Weaken to Join]
\label{thm:weakenjoin}
Let $F \subseteq \lang$ be  upward-closed w.r.t. $\sqsubseteq$,
$R_F$ its representation ($R_F = \min{}_\sqsubseteq \{c(\phi) \mid \phi \in F\}$), and $\state$ a state.
The representation of $F \join \alpha(\{\state\})$
is given by $\weaken{\lang}{R_F}{\state} = \min{}_\sqsubseteq \bigcup_{\phi \in R_F} \weaken{\lang}{\phi}{\state}$.
\end{theorem}

\begin{corollary}[Weaken for a Set of States]
\label{thm:weakenjoinset}
Let $F \subseteq \lang$ be upward-closed w.r.t. $\sqsubseteq$,
$R_F$ its representation, and $\state_1,\ldots,\state_n$ states.
The representation of $F \join \alpha(\{\state_1,\ldots,\state_n\})$
is given by $\weaken{\lang}{\weaken{\lang}{\cdots \weaken{\lang}{\weaken{\lang}{R_F}{\state_1}}{\state_2}}{\cdots \state_{n-1}}}{\state_n}$.
\end{corollary}

\begin{corollary}[Abstraction of a Set of States]
The representation of \break $\alpha(\{s_1,\ldots,s_n\})$ is given by $\weaken{\lang}{\weaken{\lang}{\cdots \weaken{\lang}{\weaken{\lang}{\{\bot_\lang\}}{\state_1}}{\state_2}}{\cdots \state_{n-1}}}{\state_n}$.
\end{corollary}

\Cref{thm:weakenjoin} and \Cref{thm:weakenjoinset} show that weakening of a single formula can be lifted to compute join between an upward-closed set of formulas (represented using its minimal elements w.r.t. $\sqsubseteq$)
and the abstraction of one or more states.

Next, we observe that we can implement $\weaken{\lang}{R}{\state}$ 
by
\begin{inparaenum}[(i)]
\item focusing only on formulas that actually need weakening, i.e., formulas in $R$ that are not satisfied by $\state$, without iterating over formulas that $\state$ satisfies; and
\item leveraging the $\leq_{\lang}$ total order to accumulate the set of minimal elements more efficiently.
\end{inparaenum}

\begin{algorithm}[t]
    \caption{In-place Weaken for $\lset{\lang}$ \label{alg:weaken-set}}
    \KwIn{An antichain of canonical $\lang$-formulas $R$ stored in the $\lset{\lang}$ data structure
      and a state $\state \in \states$
    }
    \KwOut{$R$ modified in place to store $\weaken{\lang}{R}{\state}$}
    
    \BlankLine
    $U$ :=  $\unsat{R}{\state}$\;
    \lFor{$\phi \in U$}{%
        $R$.remove($\phi$)%
    }
    $W$ := $\bigcup_{\phi \in U} \weaken{\lang}{\phi}{\state}$\;
    \For{$\phi \in W$ sorted by $\leq_\lang$}{
        \lIf{$\subsuming{R}{\phi} = \emptyset$}{%
            $R$.insert($\phi$)%
        }
    }
\end{algorithm}

\Cref{alg:weaken-set} presents our implementation of $\weaken{\lang}{R}{\state}$ for an antichain $R$ of canonical formulas and a state $\state$. It updates $R$ to $\weaken{\lang}{R}{\state}$ in place,
which is useful for computing an abstraction of a set of states (\Cref{thm:weakenjoinset}) or even for fixpoint computation (\Cref{sec:eval}).
The algorithm uses a data structure $\lset{\lang}$ (whose implementation is explained in \Cref{sec:datastructure}) that stores a set of canonical $\lang$-formulas and supports two efficient filters:
one for formulas that are not satisfied by a given state $\state$, denoted by $\unsat{R}{\state}$;
and one for formulas that subsume a given formula $\phi$, denoted by $\subsuming{R}{\phi}$. Formally:
$\unsat{R}{s} = \{\psi\in R \mid \state\not\models\psi\}$ and
$\subsuming{R}{\phi} = \{\psi\in R \mid \phi\sqsupseteq \psi\}$.

To weaken $R$, \Cref{alg:weaken-set} first identifies all formulas that need weakening using the $\unsat{R}{\state}$ filter.
It then removes these formulas, weakens them, and adds the weakened formulas back to the set, while filtering out formulas that are not $\sqsubseteq_{\lang}$-minimal.
For the minimality filtering, we leverage $\leq_{\lang}$ to ensure that if $\phi \sqsubseteq_{\lang} \psi$ then $\phi$ is added before $\psi$.
As a result, when inserting a formula $\phi$ we only need to check that it is not already subsumed by another formula in the set,
which is done by checking if $\subsuming{R}{\phi}$ is empty\footnote{While the implementation of the weaken operator only checks the emptiness of $\subsuming{R}{\phi}$, the full set is used in the recursive implementation of $\subsuming{R}{\phi}$ (\Cref{sec:datastructure}).}.
Importantly, a formula $\phi \in R \setminus \unsat{R}{\state}$ cannot be subsumed by a formula from $\weaken{\lang}{\psi}{\state}$ for $\psi \in \unsat{R}{\state}$. (If we assume the contrary we easily get that $\psi \sqsubseteq \phi$, contradicting the fact that $R$ is an antichain.)

\subsection{Design Consideration and Tradeoffs}
\label{sec:design}

We are now in a position to discuss the tradeoffs and considerations that arise in our framework in the design of languages and their subsumption relations, explaining the design choices behind \Cref{def:lang,def:subsumption}.

There is a tradeoff between the precision of the subsumption relation $\sqsubseteq_\lang$ and the complexity of implementing the weaken operator $\weakenop_\lang$.
From a representation perspective, a more precise $\sqsubseteq_\lang$ is desirable (i.e., relating more formulas), since it means that the upward closure $\up{\{\phi\}}$ of a formula $\phi$ is larger, and (upward-closed) sets of formulas can be represented using less minimal formulas.
On the other hand, when $\up{\{\phi\}}$ is larger, computing $\weaken{\lang}{\phi}{\state}$ is generally more complicated.
As an extreme case, if $\sqsubseteq_\lang$ is trivial (i.e., a formula only subsumes itself),
we get no pruning in the representation,
but computing $\weaken{\lang}{\phi}{\state}$ is very easy, since it is either $\{\phi\}$ or $\emptyset$.
As another example, compare $\orAB{\lang}{\lang}$ with $\orA{2}{\lang}$.
The subsumption relation of $\orAB{\lang}{\lang}$ is a pointwise extension,
while that of $\orA{2}{\lang}$ allows swapping the two formulas, which is more precise.
(E.g., $\bigvee \seq{\phi, \psi} \sqsubseteq_{\orA{2}{\lang}} \bigvee \seq{\psi, \phi}$ always holds but
we might have $\phi \vee \psi \not\sqsubseteq_{\orAB{\lang}{\lang}} \psi \vee \phi$.)
Accordingly, weakening of ${\orA{2}{\lang}}$-formulas is slightly more involved.

As opposed to reordering of disjuncts, $\sqsubseteq_{\orA{k}{\lang}}$ does not allow multiple disjuncts to subsume the same one, e.g., $\bigvee \seq{\phi, \psi} \not\sqsubseteq_{\orA{k}{\lang}} \bigvee \seq{\psi}$ even if $\phi\sqsubseteq_\lang\psi$ (recall that the mapping between disjuncts must be injective).
This choice makes the computation of $\weakenop_{\orA{k}{\lang}}$ simpler, as it only needs to consider individually weakening each disjunct or adding a new one, but not merging of disjuncts (to ``make space'' for a new disjunct).
For example, when computing $\weaken{\orA{2}{\lang}}{\bigvee \seq{\phi_1, \phi_2}}{\state}$,
we do not have to consider formulas of the form $\bigvee\seq{\phi, \psi}$ where $\state \models \psi$ and $\phi_1,\phi_2 \sqsubseteq_\lang \phi$, which we would need to include if the mapping was not required to be injective.
One seemingly undesirable consequence of the injectivity requirement is that canonical formulas may contain redundant disjuncts, e.g., $\bigvee \seq{\phi, \psi}$ when $\phi \sqsubseteq \psi$ (or even $\bigvee \seq{\phi, \phi}$). However, when formulas are obtained by iterative weakening, as in the computation of the representation of $\alpha(S)$ for a set of concrete states $S$, formulas with such redundancies will be eliminated as they are always subsumed by a canonical formula without redundancies.

Our design of bounded first-order languages uses bounded disjunction but unbounded conjunction.
The reason is that we obtain formulas by weakening other formulas, starting from $\bot_\lang$.
In this scenario, bounding the size of conjunctions would have replaced one conjunction by all of its subsets smaller than the bound, causing an exponential blowup in the number of formulas, without contributing much to generalization.
On the other hand, bounding the size of disjunctions yields generalization without blowing up the number of formulas (in fact, it reduces the number of formulas compared to unbounded disjunction).

\section{Data Structure for Sets of Formulas}
\label{sec:datastructure}

The implementation of $\weaken{\lang}{R}{\state}$ presented in \Cref{alg:weaken-set} uses the filters $\unsat{R}{\state}$ and $\subsuming{R}{\phi}$. Since the sets
may be very large,
a naive implementation that iterates over $R$ to find formulas that are not satisfied by $\state$ ($\unsat{R}{\state}$) or formulas that subsume $\phi$ ($\subsuming{R}{\phi}$) may become inefficient.
We therefore introduce a data structure for bounded first-order languages, which we call $\lset{\lang}$, that stores a set of canonical $\lang$-formulas $R$ (not necessarily an antichain), and implements 
$\unsat{R}{\state}$ and $\subsuming{R}{\phi}$ without iterating over all formulas in $R$.
The key idea is to define the $\lset{\lang}$ data structure recursively, following the structure of $\lang$,
and to use auxiliary data to implement the $\unsat{R}{\state}$ and $\subsuming{R}{\phi}$ filters more efficiently.

For example, to implement $\lset{\orAB{\lang_1}{\lang_2}}$,
we store a set of $\orAB{\lang_1}{\lang_2}$-formulas
and two auxiliary data fields: an LSet 
$L:\lset{\lang_1}$ and a map
$M:\stdmap{\lang_1}{\lset{\lang_2}}$.
We maintain the invariant that
$\phi_1\vee\phi_2$ is in the set iff $\phi_2\in M[\phi_1]$, and that $L$ contains the same $\lang_1$-formulas as the keys of $M$.
Then, to find formulas that are not satisfied by a state $\state$, i.e., formulas where both disjuncts are not satisfied by $\state$, we first query $L$ to find $\phi_1$'s that are not satisfied by $\state$,
and for each such $\phi_1$ we query the LSet $M[\phi_1]$ to find $\phi_2$'s that are not satisfied by $\state$.
Implementing the subsumption filter follows a similar logic.

Our implementation of $\lset{\orA{k}{\lang}}$ uses a trie data structure that generalizes the binary case.
Each edge is labeled by an $\lang$-formula, and each node represents an $\orA{k}{\lang}$-formula that is
the disjunction of the edge labels along the path from the root to the node.
The outgoing edges of each node are stored using an $\lset{\lang}$ that can be used to filter only the edges whose label is not satisfied by a given state, or subsumes a given formula. Then, the $\unsat{R}{\state}$ and $\subsuming{R}{\bigvee\tupphi}$ filters are implemented by recursive traversals of the tree that only traverse filtered edges.

The recursive implementation  for the other language constructors is simpler,
and follows a similar intuition to that of the cases presented above.
The base case $\lset{\langa}$ is implemented without any auxiliary data using straightforward iteration.
The full details of the $\lset{\lang}$ data structure appear
\ifextended
in \Cref{app:lset}.
\else
in~\cite{extended-version}.
\fi

\section{Implementation and Evaluation}
\label{sec:eval}

To evaluate our abstract domain implementation, we used it to
implement a symbolic abstraction~\cite{symbolic-thakur,symabs2004} algorithm that computes the least fixpoint of the best abstract transformer of a transition system.
We evaluated our implementation on 19 distributed protocols commonly used as benchmarks in safety verification and obtained promising results.

\subsection{Implementation}

We implemented our abstract domain and the symbolic abstraction algorithm in \flyvy,%
\footnote{\flyvy's code is available at \href{https://github.com/flyvy-verifier/flyvy}{https://github.com/flyvy-verifier/flyvy}.}
an open-source verification tool written in Rust, whose implementation leverages parallelism and the optimizations detailed below.
The implementation and benchmarks used, as well as the log files and raw results from the experiments reported, are publicly available in this paper's artifact~\cite{artifact}.

Our implementation
receives as input (i)~a first-order transition system $(\init, \tr)$ over signature $\signature$, where $\init$ is a closed first-order formula over $\signature$ specifying the initial states and $\tr$ is a closed first-order formula over two copies of $\signature$ specifying the transitions, and (ii)~a specification of a bounded first-order language $\lang$ over $\signature$ that defines the abstract domain $\powerset{\lang}$.
The reachable states of the system are the least fixpoint of a concrete transformer $\trans: \powerset{\states} \to \powerset{\states}$ given by
$ \trans(S) = \{\state' \in \states \mid \state' \models \init \lor \exists \state \in S. \ \seq{\state, \state'} \models \tr\}$, where $\seq{\state, \state'} \models \tr$ indicates that
the pair of states satisfies the two-vocabulary formula $\tr$, i.e.,
that $\state'$ is a successor of $\state$ w.r.t the transition relation defined by $\tau$.
For more details on this style of modeling distributed systems in first-order logic, see~\cite{oded-phd,paxos-made-epr,ivy-original}.

The Galois connection $(\alpha, \gamma)$ between $\powerset{\states}$ and $\powerset{\lang}$
induces a \emph{best abstract transformer} $\abs{\trans}: \powerset{\lang} \to \powerset{\lang}$ defined by $\abs{\trans} = \alpha \circ \trans \circ \gamma$.
Any fixpoint of $\abs{\trans}$, i.e., a set $F \subseteq \lang$ such that $\abs{\trans}(F) = F$, is an \emph{inductive invariant} of $(\init,\tr)$ (when sets are interpreted conjunctively), and the least fixpoint, $\lfp \abs{\trans}$, is the strongest inductive invariant in $\lang$.
The strongest inductive invariant is useful for verifying safety properties of the system,
or showing that they cannot be proven in $\lang$ (if the strongest inductive invariant in $\lang$ cannot prove safety, neither can any other inductive invariant expressible in $\lang$).

Symbolic abstraction computes $\lfp \abs{\trans}$ without computing $\abs{\trans}$ explicitly: beginning with $F= \lang$ (the least element in $\powerset{\lang}$), and as long as $F \neq \abs{\trans}(F)$, a \emph{counterexample to induction} (CTI) of $F$ is sampled, i.e., a state
$\state' \not \models \bigwedge F$ that is either an initial state or the successor of a state $s$ with $\state \models \bigwedge F$,
and $F$ is updated to $F \join \alpha(\{\state'\})$.
Our implementation uses the representation $R_F$ and \Cref{alg:weaken-set} to compute the
\ifextended
join (more details in \Cref{sec:sym-abs-alg}).
\else
join (more details in~\cite{extended-version}).
\fi
To find CTIs or determine that none exist we use SMT solvers (Z3~\cite{z3-smt-solver} and cvc5~\cite{cvc5-smt-solver}), with queries restricted to the EPR fragment (following~\cite{paxos-made-epr}),
which ensures decidability and the existence of finite counterexamples.
Solvers still struggle in some challenging benchmarks, and we employ several optimizations
\ifextended
detailed in \Cref{sec:solver-optimizations}
\else
detailed in~\cite{extended-version}
\fi
to avoid solver timeouts.

\subsection{Experiments}
\label{sec:experiments}

To evaluate our techniques, we computed the least fixpoints (strongest inductive invariants)
of 19 distributed protocols commonly used as benchmarks in safety verification, in a language expressive enough to capture their human-written safety invariants.
We used all EPR benchmarks from~\cite{p-fol-ic3}, except for universally quantified Paxos variants.
To evaluate the utility of the LSet data structure described in \Cref{sec:datastructure},
we ran each experiment twice, once using LSet and once using a naive (but parallelized) implementation for the filters $\unsat{R}{\state}$ and $\subsuming{R}{\phi}$.

To specify the bounded first-order language for each example, we provide the tool with a quantifier prefix (using $\existsA{X}{\cdot}$, $\forallA{X}{\cdot}$, and $\existsforallA{X}{\cdot}$) composed on top of a quantifier-free bounded language that captures $k$-pDNF
(following~\cite{p-fol-ic3}).
A $k$-pDNF formula has the structure $c_1\to (c_2\vee\dots\vee c_k)$, where $c_1,c_2,\dots,c_k$ are cubes (conjunctions of literals). We specify such formulas as %
$\orAB{\orA{n}{\lang_{A_1}}}{\orA{k-1}{\andA{\omega}{\lang_{A_2}}}}$, where $k$ and $n$ are parameters, and $A_1$ and $A_2$ are sets of literals.
Inspired by~\cite{duoai},
we observe that we can restrict the variables used in $A_1$ and $A_2$ to reduce the size of the language without losing precision.%
\footnote{One of the language reductions used by~\cite{duoai} relies on an overly generalized lemma~\cite[Lemma 6]{duoai}; we confirmed this with the authors of~\cite{duoai}. We prove and use a correct (but less general) variant of this lemma,
\ifextended
see \Cref{app:language-optimizations} for details.
\else
see~\cite{extended-version} for details.
\fi}
For additional details
\ifextended
see \Cref{app:language-optimizations}.
\else
see~\cite{extended-version}.
\fi
The list of examples with their language parameters appears in \Cref{tab:fixpoint}.
For each example, we report the quantifier structure, the $k$ and $n$ parameters of the $k$-pDNF quantifier-free matrix, and the approximate size of the language $\lang$. Recall that the size of the abstract domain \nolinebreak is \nolinebreak $2^{|\lang|}$.

All experiments were performed on a 48-threaded machine with 384 GiB of RAM
(AWS's \texttt{z1d.metal})
and a three-hour time limit.
For each example we also provide runtimes of two state-of-the-art safety verification tools, DuoAI~\cite{duoai} and P-FOL-IC3~\cite{p-fol-ic3}.
Note that, unlike our technique, these tools look for \emph{some} inductive invariant proving safety, not necessarily the strongest,
but
are also given fewer explicit language constraints.
Moreover, the runtimes of DuoAI and P-FOL-IC3 are sourced from their respective papers, and reflect different architectures and time limits. Thus, the inclusion of their results is not intended as a precise comparison to our tool, but as a reference for the difficulty of the invariant inference task of each example, as evidenced by state-of-the-art techniques.

\subsection{Results}
\begingroup
\begin{table}[!ht]
\caption{
Symbolic abstraction over invariant inference benchmarks with a time limit of 3 hours (10800s).
We describe
the bounded language underlying the abstract domain of each example, including its approximate size,
and report
the runtime of our technique---with and without using LSet---along with some statistics.
For reference, we provide runtimes of two state-of-the-art safety-verification tools.
`T/O' indicates a timeout, and `N/A' indicates that the example was not reported by the respective tool.
\label{tab:fixpoint}
}
\centering
\begin{adjustbox}{width=\textwidth,keepaspectratio}
\begin{tblr}{
        columns = {colsep=2pt},
        rows = {rowsep=0.5pt},
		cell{-}{-} = {c},
	    cell{1}{1,6-10} = {r=2}{},
        cell{1}{2} = {c=4}{},
        cell{1}{11} = {c=2}{},
	    cell{3,5,7,9,11,13,15,17,19,21,23,25,27,29,31,33,35,37,39}{1-5,9,11,12} = {r=2}{},
        hline{1,3,21,41} = {},
        hline{5,7,9,11,13,15,17,19,23,25,27,29,31,33,35,37,39} = {dashed},
        hline{4,6,8,10,12,14,16,18,20,22,24,26,28,30,32,34,36,38,40} = {dotted},
        vline{1,2,6,11,13} = {}
}
\textbf{Example} &
\textbf{Language} & & & &
\makecell{\textbf{Runtime}\\(sec)} &
LSet &
\% in $\mathcal{W}$ &
\ifrange Lfp. Size \else \makecell{Lfp.\\Size} \fi &
\ifrange Max. Size \else \makecell{Max.\\Size} \fi &
\textbf{Safety} (sec) & \\
& quant. & $k$ & $n$ & size & & & & & & P-FOL-IC3 & DuoAI \\
lockserv & $\forall^2$ & 1 & 3 & $10^4$ &
$\mathbf{0.4 \pm 0.1}$ & \checkmark & $6 \ifrange\pm 1\fi\ \%$ & 12 & $28 \ifrange\pm 7\fi$&
19 & 1.9 \\
& & & & &
$0.5 \pm 0.1$ & -- & $6 \ifrange\pm 1\fi\ \%$ & & $29 \ifrange\pm 3\fi$ \\
\makecell{toy-consensus-\\forall} & $\forall^3$ & 1 & 3 & $10^3$ &
$\mathbf{0.2 \pm 0.0}$ & \checkmark & $9 \ifrange\pm 2\fi\ \%$ & 5 & $18 \ifrange\pm 5\fi$ &
 4 & 1.9 \\
& & & & &
$0.2 \pm 0.0$ & -- & $7 \ifrange\pm 1\fi\ \%$ & & $18 \ifrange\pm 4\fi$ \\
ring-id & $\forall^3$ & 1 & 3 & $10^5$ &
$\mathbf{1.6 \pm 0.1}$ & \checkmark & $16 \ifrange\pm 1\fi\ \%$ & 97 & $182 \ifrange\pm 22\fi$ &
7 & 3.5 \\
& & & & &
$1.9 \pm 0.1$ & -- & $20 \ifrange\pm 1\fi\ \%$ & & $189 \ifrange\pm 22\fi$ \\
sharded-kv & $\forall^5$ & 1 & 3 & $10^4$ &
$\mathbf{0.5 \pm 0.1}$ & \checkmark & $8 \ifrange\pm 0\fi\ \%$ & 20 & $26 \ifrange\pm 2\fi$ &
8 & 1.9 \\
& & & & &
$0.5 \pm 0.0$ & -- & $8 \ifrange\pm 1\fi\ \%$ & & $26 \ifrange\pm 4\fi$ \\
ticket & $\forall^4$ & 1 & 5 & $10^9$ &
$\mathbf{32.6 \pm 3.3}$ & \checkmark & $43 \ifrange\pm 7\fi\ \%$ & 2621 & $8531 \ifrange\pm 119\fi$ &
23 & 23.9 \\
& & & & &
$862.2 \pm 21.9$ & -- & $97 \ifrange\pm 0\fi\ \%$ & & $8533 \ifrange\pm 121\fi$ \\
\makecell{learning-\\switch} & $\forall^4$ & 1 & 4 & $10^{11}$ &
\textbf{T/O} & \checkmark & $98\ \%$ & -- & 9576194 &
76 & 52.4 \\
& & & & &
T/O & -- & $100\ \%$ & & 5998 \\
\makecell{consensus-\\wo-decide} & $\forall^3$ & 1 & 3 & $10^6$ &
$\mathbf{3.0 \pm 0.2}$ & \checkmark & $19 \ifrange\pm 1\fi\ \%$ & 41 & $717 \ifrange\pm 109\fi$ &
50 & 3.9 \\
& & & & &
$4.0 \pm 0.6$ & -- & $38 \ifrange\pm 4\fi\ \%$ & & $724 \ifrange\pm 43\fi$ \\
\makecell{consensus-\\forall} & $\forall^4$ & 1 & 3 & $10^6$ &
$\mathbf{3.5 \pm 0.4}$ & \checkmark & $21 \ifrange\pm 2\fi\ \%$ & 51 & $740 \ifrange\pm 114\fi$ &
1980 & 11.9 \\
& & & & &
$5.1 \pm 0.9$ & -- & $40 \ifrange\pm 6\fi\ \%$ & & $708 \ifrange\pm 82\fi$ \\
cache & $\forall^6$ & 1 & 5 & $10^{11}$ &
$\mathbf{4029.4 \pm 220.7}$ & \checkmark & $30 \ifrange\pm 2\fi\ \%$ & 106348 & $271255 \ifrange\pm 13081\fi$ &
2492 & N/A \\
& & & & &
T/O & -- & $100 \pm 0\ \%$ & & $19183 \ifrange\pm 9466\fi$ \\
\makecell{sharded-kv-\\no-lost-keys} & $\forall^1 (\existsforall)^2$ & 1 & 2 & $10^2$ &
$\mathbf{0.3 \pm 0.0}$ & \checkmark & $3 \ifrange\pm 0\fi\ \%$ & 4 & $4 \ifrange\pm 0\fi$ &
4 & 2.1 \\
& & & & &
$0.3 \pm 0.0$ & -- & $2 \ifrange\pm 0\fi\ \%$ & & $4 \ifrange\pm 0\fi$ \\
\makecell{toy-consensus-\\epr} & $\forall^2 (\existsforall)^1 \forall^1$ & 1 & 3 & $10^4$ &
$\mathbf{0.3 \pm 0.0}$ & \checkmark & $9 \ifrange\pm 1\fi\ \%$ & 5 & $18 \ifrange\pm 4\fi$ &
4 & 2.6 \\
& & & & &
$0.3 \pm 0.0$ & -- & $7 \ifrange\pm 1\fi\ \%$ & & $19 \ifrange\pm 3\fi$ \\
\makecell{consensus-\\epr} & $(\existsforall)^1 \forall^4$ & 1 & 3 & $10^6$ &
$\mathbf{5.1 \pm 0.7}$ & \checkmark & $17 \ifrange\pm 2\fi\ \%$ & 51 & $800 \ifrange\pm 137\fi$ &
37 & 4.8 \\
& & & & &
$8.8 \pm 1.7$ & -- & $46 \ifrange\pm 8\fi\ \%$ & & $783 \ifrange\pm 88\fi$ \\
\makecell{client-\\server-ae} & $\forall^2 (\existsforall)^1$ & 2 & 1 & $10^3$ &
$\mathbf{0.2 \pm 0.0}$ & \checkmark & $4 \ifrange\pm 1\fi\ \%$ & 2 & $5 \ifrange\pm 0\fi$ &
4 & 1.5 \\
& & & & &
$0.2 \pm 0.1$ & -- & $3 \ifrange\pm 1\fi\ \%$ & & $5 \ifrange\pm 0\fi$ \\
paxos-epr & $\forall^4 (\existsforall)^2$ & 2 & 3 & $10^{11}$ &
$\mathbf{621.5 \pm 246.8}$ & \checkmark & $1 \ifrange\pm 1\fi\ \%$ & 1438 & $1693 \ifrange\pm 203\fi$ &
920 & 60.4 \\
& & & & &
$789.6 \pm 285.5$ & -- & $11 \ifrange\pm 3\fi\ \%$ & & $1737 \ifrange\pm 168\fi$ \\
\makecell{flexible-\\paxos-epr} & $\forall^4 (\existsforall)^2$ & 2 & 3 & $10^{11}$ &
$\mathbf{166.7 \pm 29.3}$ & \checkmark & $4 \ifrange\pm 1\fi\ \%$ & 964 & $1622 \ifrange\pm 177\fi$ &
418 & 78.7 \\
& & & & &
$235.6 \pm 31.7$ & -- & $35 \ifrange\pm 8\fi\ \%$ & & $1575 \ifrange\pm 196\fi$ \\
\makecell{multi-\\paxos-epr} & $\forall^5 (\existsforall)^3$ & 2 & 3 & $10^{30}$ &
\textbf{T/O} & \checkmark & $2\ \%$ & -- & 27508 &
4272 & 1549 \\
& & & & &
T/O & -- & $100\ \%$ & & 6400 \\
\makecell{fast-\\paxos-epr} & $\forall^4 (\existsforall)^3$ & 2 & 4 & $10^{14}$ &
\textbf{T/O} & \checkmark & $1\ \%$ & -- & 16290 &
9630 & 26979 \\
& & & & &
T/O & -- & $99\ \%$ & & 13683 \\
\makecell{stoppable-\\paxos-epr} & $\forall^7 (\existsforall)^3$ & 2 & 5 & $10^{155}$ &
\textbf{T/O} & \checkmark & $100\ \%$ & -- & 37529 &
$>$18297 & 4051 \\
& & & & &
T/O & -- & $100\ \%$ & & 3331 \\
\makecell{vertical-\\paxos-epr} & $\forall^4 (\existsforall)^3$ & 3 & 5 & $10^{54}$ &
\textbf{T/O} & \checkmark & $100\ \%$ & -- & 112990 &
T/O & T/O \\
& & & & &
T/O & -- & $100\ \%$ & & 2576
\end{tblr}
\end{adjustbox}
\end{table}
\endgroup

The results of the symbolic abstraction computation are presented in \Cref{tab:fixpoint}.
For each experiment we report the runtime of our tool and the following statistics: the percentage of time spent weakening formulas (as opposed to searching for CTIs), the number of formulas in the representation of the fixpoint (if reached), and the maximal number of formulas in the representation of an abstract element throughout the run.
Each experiment was run five times, unless it timed out, in which case it was run only once.
We aggregate the results of each statistic across multiple runs as $median \pm deviation$, where $deviation$ is the maximal distance between the median value and the value of the statistic in any given run.

For simple examples, the fixpoint computation terminates very quickly, often faster than the other tools, and maintains only tens or hundreds of formulas throughout its run.
Some of the larger examples, such as \texttt{ticket}, \texttt{paxos-epr}, \texttt{flexible-paxos-epr}, and \texttt{cache} also terminate after similar times to the other tools. In fact, this is the first work to compute least fixpoints for any Paxos variant or \texttt{cache}. (DuoAI, for instance, has a component that attempts to compute a precise fixpoint, but~\cite{duoai} reports that it times out on all Paxos variants.)

Unsurprisingly, there is a significant gap between the runtimes of examples with and without quantifier alternation, mostly due to the time spent in SMT solvers. For example, in \texttt{ticket} we spend about $43\%$ of the runtime performing weakenings, but this percentage drops to $1\%$ and $4\%$ for \texttt{paxos-epr} and \texttt{flexible-paxos-epr}, respectively. This causes the runtime of \texttt{paxos-epr} to exceed that of \texttt{ticket} by more than an order of magnitude, although its fixpoint computation considers fewer formulas and actually spends less time weakening.
Similarly, in \texttt{cache} we manage to prove a fixpoint of a hundred thousand formulas in about an hour and spend a third of it weakening formulas, while \texttt{multi-paxos-epr} and \texttt{fast-paxos-epr} time out, although they consider far fewer formulas and spend a negligible amount of time weakening.

Next, we observe that the use of LSet significantly reduces time spent in weakening, leading to more than an order of magnitude difference even in moderate examples, e.g., \texttt{ticket} and \texttt{paxos-epr}. In terms of the \emph{total} fixpoint computation time, in examples where the runtime is small or dominated by the SMT solvers, the effect might be negligible, but otherwise the speedup is significant. For example, \texttt{cache} is not solved within the 3-hour limit with a naive data structure; it gets stuck after reaching $\sim$ 20,000 formulas in the abstraction, whereas using LSet it is solved in about an hour while handling more than ten times the number of formulas. Similarly, in the two unsolved examples where SMT calls seem to be the bottleneck (\texttt{multi-paxos-epr} and \texttt{fast-paxos-epr}), using a naive data structure causes weakening to become the bottleneck and time out.

Finally, the remaining timeouts, \texttt{learning-switch}, \texttt{stoppable-paxos-epr}, and \texttt{vertical-paxos-epr}, are the only examples where the weakening process itself is the bottleneck. These are cases where the language induced by the human-written invariant, using the constraining parameters of bounded languages, create a inefficient weakening process. The cause for this is either a profusion of literals in the basis language ($>$ 600 in \texttt{learning-switch} and \texttt{stoppable-paxos-epr}, less than 200 in all other examples), or a very expressive language (e.g., \texttt{vertical-paxos-epr} uses 3-pDNF, whereas all other examples use 1- and 2-pDNF). For these examples, it might be necessary to restrict the languages in additional ways, e.g., as was done in~\cite{duoai}.
Our experience, however, is that the more significant bottleneck for computing least fixpoints for the most complicated examples is the SMT queries.

\section{Related Work}
\label{sec:related}
Many recent works tackle invariant inference in first-order logic~\cite{ic3po,distai,duoai,%
	updr,swiss,%
	ic3po-nfm,%
	i4,p-fol-ic3,%
	induction-duality%
}. These works are all property-guided and employ sophisticated heuristics to guide the search for invariants. Of these works, the most closely related to ours are~\cite{distai,duoai}. DistAI~\cite{distai} is restricted to universally quantified invariants, while DuoAI~\cite{duoai} infers invariants with quantifier alternations. 
DuoAI defines a ``minimum implication graph'' enumerating all formulas in a first-order logical language, whose transitive closure can be understood as a specific subsumption relation, and where replacing a node with its successors can be understood as a form of weakening. DuoAI's ``top-down refinement'' precisely computes the strongest invariant in the logical domain. However, this computation does not scale to complex examples such as all Paxos variants, in which case 
``bottom-up refinement'' is used---a property-guided process that does not compute the strongest invariant.
Our approach based on a generic subsumption relation is both more principled and more scalable,
as it succeeds in computing the least fixpoint for some Paxos variants.

Another work concerning a least-fixpoint in a logical domain is \cite{mcmillan-k-clause},
which computes the set of propositional clauses up to length $k$ implied by a given formula, minimized by the subsumption relation $\sqsubseteq=\subseteq$; a trie-based data structure is used to maintain the formulas, weaken them, and check subsumption of a formula by the entire set.
Both that data structure and $\lset{\orA{k}{\cdot}}$ bear similarity to UBTrees \cite{ubtrees}, also employed in \cite{klee}, which store sets and implement filters for subsets and supersets. However, while UBTrees and LSets always maintain ordered tree paths, these are unordered in \cite{mcmillan-k-clause}, which allows \cite{mcmillan-k-clause} to perform weakening directly on the data structure, whereas we need to remove the unsatisfied disjunctions, weaken, and insert them. On the other hand, this makes filtering for subsets in UBTrees and LSets more efficient. Also note that LSet is more general than both, since it supports a more general subsumption relation.

\section{Conclusion}
\label{sec:conclusion}

We have developed key algorithms and data structures for working with a logical abstract domain of quantified first-order formulas. Our fundamental idea is using a well-defined subsumption relation and a \emph{weaken} operator induced by it. This idea makes the abstract domain feasible, and it is also extensible: while we explored one possible subsumption relation and its associated weaken operator, future work may explore others, representing different tradeoffs between pruning and weakening. We demonstrated the feasibility of our approach by computing the least abstract fixpoint for several distributed protocols modeled in first-order logic---a challenging application domain where previously only property-directed heuristics have been successful.
For some of the examples in our evaluation, the computation still times out. 
In some of these cases, SMT queries (for computing CTIs) become the bottleneck.
Dealing with this bottleneck is an orthogonal problem that we leave for future work.
For the examples with the largest logical languages, abstract domain operations remain the bottleneck, and future work may either scale the abstract domain implementation to such languages or explore combinations with property-directed approaches.

\begin{credits}
\subsubsection{\ackname}
We thank Alex Fischman and James R. Wilcox for their contributions to the \flyvy\ verification tool. We thank Raz Lotan, Kenneth McMillan, and the anonymous reviewers for their helpful and insightful comments.

The research leading to these results has received funding from the
European Research Council under the European Union's Horizon 2020 research and innovation programme (grant agreement No [759102-SVIS]).
This research was partially supported by the Israeli Science Foundation (ISF) grant No.\ 2117/23.
\end{credits}

\clearpage

\bibliographystyle{splncs04}
\bibliography{refs}

\ifextended
\clearpage

\appendix
\section{Running Example}
\label{app:running-example}

\subsection{Bounded First-Order Language}
\label{app:example-language}

Consider the signature $\signature$ with one unary predicate $p$, and variables $V=\{x,y\}$. These induce the set of first-order literals
$ A = \{ p(x),\neg p(x), p(y),\neg p(y) \} $,
which can be used to define, for example, the bounded language $\lang = \forallA{\{x,y\}}{\orA{2}{\lang_A}}$.
Formulas in this language are universally quantified (homogeneous) disjunctions of at most two literals. For instance, $\lang$ includes 
\begin{itemize}
	\item $\forall\{x,y\}.\bigvee\emptytup$, which is also $\bot_\lang$, 
	\item $\forall\{x,y\}.\bigvee\seq{p(x)}$,
	\item $\forall\{x,y\}.\bigvee\seq{p(y), p(x)}$,
	\item $\forall\{x,y\}.\bigvee\seq{p(x), \neg p(y)}$,
	\item $\forall\{x,y\}.\bigvee\seq{p(x), p(x)}$,
	\item etc.
\end{itemize}

\subsection{Subsumption}
\label{app:example-subsumption}

The language $\lang = \forallA{\{x,y\}}{\orA{2}{\lang_A}}$ defined in \Cref{app:example-language} is composed, hierarchically, using three bounded first-order languages: $\lang_A$, $\orA{2}{\lang_A}$, and $\forallA{\{x,y\}}{\orA{2}{\lang_A}}$. Each is equipped with its own subsumption relation as defined inductively in \Cref{def:subsumption}. Observe that $p(x)\not\sqsubseteq_{\lang_A} p(y)$ and $\bigvee\seq{p(x)} \not\sqsubseteq_{\orA{2}{\lang_A}} \bigvee\seq{p(y)}$, but
\[ \forall\{x,y\}.\bigvee\seq{p(x)} \quad \sqsubseteq_{\lang} \quad \forall\{x,y\}.\bigvee\seq{p(y)}, \]
because by permuting $x$ and $y$ we can infer the above from $\bigvee\seq{p(x)}\sqsubseteq\bigvee\seq{p(x)}$. Thus, one source of subsumption is due to permutation of variables in ways that do not change the semantics of a formula. Another is due to reordering subformulas in a homogeneous connective, for example,
\[ \bigvee\seq{p(x),p(y)} \quad\sqsubseteq_{\orA{2}{\lang_A}}\quad \bigvee\seq{p(y),p(x)}. \]

What the above demonstrates is that subsumption approximates semantic entailment by utilizing the local semantic characteristics of each language. It is indeed not the case that $p(x)\models p(y)$, but when both are quantified universally, or when both are put in a disjunction and then reordered, we can infer something useful about their semantics using simple facts like $p(x)\models p(x)$ and $p(y)\models p(y)$, which are trivial and thus assumed by the basis subsumption of $\lang_A$.

\subsection{Canonicalization}
\label{app:example-canonicalization}
In order to have a canonicalization w.r.t $\equiv_{\sqsubseteq_\lang}$ for $\lang = \forallA{\{x,y\}}{\orA{2}{\lang_A}}$ defined in \Cref{app:example-language}, we need to provide an arbitrary total order $\leq_{\lang_A}$. Say we choose
\[ p(x) <_{\lang_A} \neg p(x) <_{\lang_A} p(y) <_{\lang_A} \neg p(y), \]
and recall that $\bot\in\lang_A$ is least.
\Cref{def:canonicalization}, together with \Cref{thm:canonicalization}, show that the above induces a canonicalization $c_\lang$, as well as a total order $\leq_\lang$ over canonical $\lang$-formulas which extends $\sqsubseteq_\lang$.

To see how the canonicalization $c_\lang$ eliminates subsumption equivalence, consider the two subsumption-equivalent formulas presented in \Cref{app:example-subsumption}:
\[ \forall\{x,y\}.\bigvee\seq{p(x)} \quad \equiv_{\sqsubseteq_{\lang}} \quad \forall\{x,y\}.\bigvee\seq{p(y)}. \]
The canonicalization of both is $\forall\{x,y\}.\bigvee\seq{p(x)}$, since $c_\lang$ selects the minimal canonical body after applying all variable permutations over $x,y$ (\Cref{def:canonicalization}), and $p(x)<_{\lang_A}p(y)$.
Note that $\bigvee\seq{p(x)}$ and $\bigvee\seq{p(y)}$ are both canonical, but adding quantifiers merges the two formulas into the same subsumption-equivalence class, necessarily making the quantified version of one of them non-canonical.

Similarly, at the level of $\orA{2}{\lang_A}$,
\[ \bigvee\seq{p(x),p(y)} \quad\equiv_{\sqsubseteq_{\orA{2}{\lang_A}}}\quad \bigvee\seq{p(y),p(x)}, \]
and canonicalization is achieved by sorting the sequences of literals, so both are canonicalized as $\bigvee\seq{p(x),p(y)}$.
Interestingly, observe that for the quantified formula
$\forall\{x,y\}. \bigvee\seq{p(x),p(y)}$,
any variable permutation applied the quantifier-free body, after canonicalization w.r.t $\orA{2}{\lang_A}$, i.e., sorting, yields $\bigvee\seq{p(x),p(y)}$. Thus, for the subsumption-equivalence class of this formula, canonicalization w.r.t $\orA{2}{\lang_A}$ also eliminates any subsumption-equivalence due to quantification. This is not entirely surprising, since in this case, permuting the variables and permuting the literals in the disjunction has the same effect.

\subsection{Weakening}
\label{app:example-weaken}

As an example, let us weaken the bottom element $\bot_\lang$ for $\lang = \forallA{\{x,y\}}{\orA{2}{\lang_A}}$, i.e., $\forall \{x,y\}.\bigvee \epsilon$, with the state $s=((\univ,\interp),\mu)$ where $\univ=\{a,b\}$, $\interp(p)=\{a,b\}$, and $\mu$ is an arbitrary assignment. At the topmost level, the weakening procedure for $\lang=\forallA{\{x,y\}}{\orA{2}{\lang_A}}$, detailed in \Cref{thm:weaken-implementation}, requires us to first compute
\[ \Omega_{\phi_0}\left(\{\state_1,\ldots,\state_n\}\right) =
\{\phi_n \mid
\phi_1 \in \weaken{\orA{2}{\lang_A}}{\phi_0}{\state_1},
\ldots,
\phi_n \in \weaken{\orA{2}{\lang_A}}{\phi_{n-1}}{\state_n}
\} \]
where $\phi_0=\bigvee\epsilon$, and $s_1,\ldots,s_n$ are all updates of $s$ with assignments to $x,y$. The above can be computed iteratively as follows:
\begin{itemize}
	\item $R_0 := \{\bigvee\epsilon\}$;
	\item $R_1 := \bigcup_{\phi\in R_0} \weaken{\orA{2}{\lang_A}}{\phi}{\state_1}$;
	\item $\cdots$
	\item $R_n := \bigcup_{\phi\in R_{n-1}} \weaken{\orA{2}{\lang_A}}{\phi}{\state_n}$.
\end{itemize}
In our case $n=4$, and all updated states satisfy $p(x)$ and $p(y)$. Therefore, $R_1 = \{\bigvee \seq{p(x)}, \bigvee \seq{p(y)}\}$, and no formula is weakened further, since both formulas in $R_1$ are satisfied by the remaining $s_2,s_3,s_4$. Thus, $R_1=\cdots=R_n$. As expected, at the end of this process we are guaranteed to get formulas satisfying all possible updates of $s$ with assignments to $x,y$.

As we saw in \Cref{app:example-canonicalization}, both of the resulting formulas are canonical; however, adding the requisite quantification results in one non-canonical formula. This demonstrates the need for additional canonicalization for $\forallA{X}{\cdot}$.

Note the savings the subsumption-based representation affords us even for this simple language. The singleton $\{\bot_\lang\}$, which we effectively weakened per \Cref{alg:weaken-set}, represents the following abstract element containing 21 formulas:
\begin{scriptsize}
\begin{align*}
	\lang = \Big\{\ 
	&\forall \{x,y\}.\bigvee \epsilon,\  
	\forall \{x,y\}.\bigvee\seq{p(x)},\ \forall \{x,y\}. \bigvee\seq{\neg p(x)},\ 
	\forall \{x,y\}.\bigvee\seq{p(y)},\ \forall \{x,y\}. \bigvee\seq{\neg p(y)}, \\
	& \forall \{x,y\}.\bigvee\seq{p(x), p(x)},\ \forall \{x,y\}. \bigvee\seq{p(x), p(y)},\ 
	\forall \{x,y\}.\bigvee\seq{p(x), \neg p(x)},\ \forall \{x,y\}. \bigvee\seq{p(x), \neg p(y)}, \\
	& \forall \{x,y\}.\bigvee\seq{\neg p(x), p(x)},\ \forall \{x,y\}. \bigvee\seq{\neg p(x), p(y)},\ 
	\forall \{x,y\}.\bigvee\seq{\neg p(x), \neg p(x)},\ \forall \{x,y\}. \bigvee\seq{\neg p(x), \neg p(y)} \\
	& \forall \{x,y\}.\bigvee\seq{p(y), p(x)},\ \forall \{x,y\}. \bigvee\seq{p(y), p(y)},\ 
	\forall \{x,y\}.\bigvee\seq{p(y), \neg p(x)},\ \forall \{x,y\}. \bigvee\seq{p(y), \neg p(y)}, \\
	& \forall \{x,y\}.\bigvee\seq{\neg p(y), p(x)},\ \forall \{x,y\}. \bigvee\seq{\neg p(y), p(y)},\ 
	\forall \{x,y\}.\bigvee\seq{\neg p(y), \neg p(x)},\ \forall \{x,y\}. \bigvee\seq{\neg p(y), \neg p(y)}
	\ \Big\}.
\end{align*}
\end{scriptsize}%
Moreover, the singleton $\{\forall\{x,y\}.\bigvee\seq{p(x)}\}$ we got from the weakening process represents the following abstract element containing 14 formulas:
\begin{scriptsize}
	\begin{align*}
		\hspace{2em}\Big\{\ 
		&\forall \{x,y\}.\bigvee\seq{p(x)},\ \forall \{x,y\}. \bigvee\seq{p(y)}, \\
		& \forall \{x,y\}.\bigvee\seq{p(x), p(x)},\ \forall \{x,y\}. \bigvee\seq{p(x), p(y)},\ 
		\forall \{x,y\}.\bigvee\seq{p(x), \neg p(x)},\ \forall \{x,y\}. \bigvee\seq{p(x), \neg p(y)}, \\
		& \forall \{x,y\}.\bigvee\seq{\neg p(x), p(x)},\ \forall \{x,y\}. \bigvee\seq{\neg p(x), p(y)}, \\
		& \forall \{x,y\}.\bigvee\seq{p(y), p(x)},\ \forall \{x,y\}. \bigvee\seq{p(y), p(y)},\ 
		\forall \{x,y\}.\bigvee\seq{p(y), \neg p(x)},\ \forall \{x,y\}. \bigvee\seq{p(y), \neg p(y)}, \\
		& \forall \{x,y\}.\bigvee\seq{\neg p(y), p(x)},\ \forall \{x,y\}. \bigvee\seq{\neg p(y), p(y)}
		\ \Big\},
	\end{align*}
\end{scriptsize}%
but throughout the weakening process we only dealt with at most two formulas.

\section{Proofs for Subsumption-Based Representation}
\label{app:proofs}

\subsection{Correctness of Subsumption}
\begin{proof}[\Cref{thm:subsumption}]
	The proof follows the inductive definition of subsumption in \Cref{def:subsumption}.
	The semantic claim, that for each $\phi,\psi\in\lang$ if $\phi\sqsubseteq\psi$ then $\phi\models\psi$, is straightforward to verify using the definition of $\phi\sqsubseteq\psi$ for each case: For the base case we have $\phi=\bot$ or $\phi=\psi$, which implies $\phi\models\psi$. For both disjunction cases, if each disjunct of $\phi$ has a disjunct of $\psi$ that it subsumes, and therefore entails, then any state satisfying $\phi$, which therefore satisfies one of its disjuncts, also satisfies some disjunct of $\psi$. For both conjunction cases, if each conjunct of $\psi$ has a conjunct of $\phi$ that subsumes it, and therefore entails it, than any state satisfying $\phi$, which therefore satisfies all of its conjuncts, must also satisfy all of the conjuncts of $\psi$. These conditions for disjunction and conjunction, respectively, are ensured by the respective subsumption relations. For the case of quantifiers, adding the same quantification to any $\phi$ and $\psi$, or quantifying $\phi$ universally and $\psi$ existentially, as well as permuting quantified variables, all maintain entailment, so the semantic property of subsumption can be directly lifted from the induction hypothesis.
	
	Next, we prove by induction that $\sqsubseteq_\lang$ is a preorder for all $\lang$; i.e., that it is reflexive and transitive. For the base case of $\sqsubseteq_{\lang_A}$ the claim clearly holds. For $\lang_1\circ\lang_2$ where $\circ\in\{\vee,\wedge\}$, $\sqsubseteq_{\lang_1\circ\lang_2}$ is the standard preorder resulting from pointwise extension, which maintains reflexivity and transitivity. For $\orA{k}{\lang}$, reflexivity is shown by choosing the mapping $m(i) = i$ and using the reflexivity of $\sqsubseteq_\lang$ (i.h.); regarding transitivity, given $\bigvee \tupphi \sqsubseteq \bigvee \tuppsi \sqsubseteq \bigvee \tup\xi$, there exist injective $m_1 : \indices{\tupphi}\to\indices{\tuppsi}$ and $m_2 : \indices{\tuppsi}\to\indices{\tup\xi}$, so $m_2\circ m_1 : \indices{\tupphi}\to\indices{\tup\xi}$ is injective and guarantees $\phi_i \sqsubseteq_\lang \psi_{m_1(i)} \sqsubseteq_\lang \xi_{(m_2\circ m_1)(i)}$, and using the transitivity of $\sqsubseteq_\lang$ (i.h.), $\phi_i \sqsubseteq_\lang \xi_{(m_2\circ m_1)(i)}$. The cases of $\andA{\omega}{\lang}$ and $\quantA{X}{\lang}$ for $Q\in\{\exists,\forall\}$ are proven in a similar manner, by choosing the identity mapping or permutation, respectively, to show reflexivity, and composing mappings or permutations to show transitivity. For $\existsforallA{X}{\lang}$, the subsumption relation extends the subsumption relations of $\quantA{X}{\lang}$ for $Q\in\{\exists,\forall\}$, and reflexivity and transitivity can easily be lifted from them.

	Lastly, the fact that $\bot_\lang\sqsubseteq_\lang \phi$ for any $\phi\in\lang$ easily follows from its definition, using a similar inductive argument over the structure of bounded languages.
\end{proof}

\subsection{Correctness of Canonicalization}

\begin{proof}[\Cref{thm:canonicalization}]
	The proof follows the mutually inductive definitions of the canonicalization and the total order in \Cref{def:canonicalization}.
	To avoid confusion, we write $c(\cdot)$, $\sqsubseteq$, and $\leq$ to refer to the canonicalization and orders of the language for which the claim is proven in the induction step (omitting the subscript).
	
	The fact that $c$ is representative, i.e., that $c(\phi)\equiv_{\sqsubseteq} \phi$ for all $\phi$, is apparent in almost all cases, since they simply canonicalize subformulas in ways which, given the induction hypothesis, clearly result in subsumption-equivalent formulas.
	The only interesting case is $\andA{\omega}{\lang}$, where
	\[ c_\andA{\omega}{\lang}(\bigwedge \tupphi) = \bigwedge \ordered{\leq_\lang}{\min{}_{\sqsubseteq_\lang} \{ c_\lang(\phi_1),\ldots,c_\lang(\phi_{\card{\tupphi}})\}}, \]
	since there might be fewer formulas in $\tupphi_c = \ordered{\leq_\lang}{\min{}_{\sqsubseteq_\lang} \{ c_\lang(\phi_1),\ldots,c_\lang(\phi_{\card{\tupphi}})\}}$ than in $\tupphi$, due to minimization, and the mappings from $\indices{\tupphi}$ to $\indices{\tupphi_c}$ and from $\indices{\tupphi_c}$ to $\indices{\tupphi}$ necessary to show mutual subsumption are not obvious. However, due to $c_\lang$ being representative and decisive (i.h.), any formula in $\tupphi$ has a formula subsuming it in $\tupphi_c$; and any formula in $\tupphi_c$ has a subsumption-equivalent formula in $\tupphi$. These facts precisely prove the existence of the required mappings.
	
	For the decisiveness of $c$, we need to show that $\phi\equiv_\sqsubseteq\psi \iff c(\phi)=c(\psi)$. We begin with the ``$\impliedby$'' direction. This is again straightforward in all cases aside from $\andA{\omega}{\lang}$. For this case, denote $\bigwedge\tupphi_c = c_\andA{\omega}{\lang}(\bigwedge \tupphi)$ and $\bigwedge\tuppsi_c = c_\andA{\omega}{\lang}(\bigwedge \tuppsi)$, and assume $\bigwedge\tupphi_c = \bigwedge\tuppsi_c$. Every formula in $\tupphi$ has a formula in $\tupphi_c$ subsuming it, which is therefore also in $\tuppsi_c$, and has a subsumption-equivalent formula in $\tuppsi$. Thus, there is a mapping $m : \indices{\tupphi}\to\indices{\tuppsi}$ showing $\bigwedge\tuppsi\sqsubseteq\bigwedge\tupphi$, and the existence of a mapping for the other direction is symmetrical.
	
	Next, we recognize that the ``$\implies$'' direction of decisiveness for $c$ is implied by its representativeness together with $\sqsubseteq$ being a partial order (i.e., an anti-symmetric preorder) over canonical formulas, since then
	\[ \phi\equiv_\sqsubseteq\psi \quad\implies\quad c(\phi)\equiv_\sqsubseteq c(\psi) \quad\implies\quad c(\phi)=c(\psi). \]
	We have already dealt with $c$  being representative, and $\sqsubseteq$ being a partial order is implied by $\leq$ being a total order over canonical formulas that extends $\sqsubseteq$, which we also need to prove. Therefore, it suffices to prove the last claim.
	
	The fact that $\leq$ is a total order can easily proven by observing that the inductive definitions either lift total orders in pointwise manners ($\orAB{\cdot}{\cdot}$, $\andAB{\cdot}{\cdot}$, $\existsA{X}{\cdot}$, $\forallA{X}{\cdot}$), in lexicographic manners ($\orA{k}{\cdot}$, $\andA{\omega}{\cdot}$), or by chaining them ($\existsforallA{X}{\cdot}$). Showing that $\leq$ extends $\sqsubseteq$ is more subtle. In the base case this holds by definition. For $\orandAB{\lang_1}{\lang_2}$ where $\circ\in\{\vee,\wedge\}$, given canonical $\phi_1\circ\phi_2 \sqsubseteq \psi_1\circ\psi_2$ it holds that $\phi_1\sqsubseteq_{\lang_1}\psi_1$ and $\phi_2\sqsubseteq_{\lang_2}\psi_2$, and therefore (i.h.) $\phi_1\leq_{\lang_1}\psi_1$ and $\phi_2\leq_{\lang_2}\psi_2$, which implies $\phi_1\circ\phi_2 \leq \psi_1\circ\psi_2$.
	
	For $\orA{k}{\lang}$, given canonical $\bigvee\tupphi \sqsubseteq \bigvee\tuppsi$, there exists an \emph{injective} $m : \indices{\tupphi}\to\indices{\tuppsi}$ such that for all $i\in \indices{\tupphi}$, $\phi_i \sqsubseteq_\lang \psi_{m(i)}$, and thus $\phi_i \leq_\lang \psi_{m(i)}$. Since $\tupphi$ and $\tuppsi$ are canonical, they are ordered by $\leq_\lang$. Consider the following cases:
	\begin{itemize}
		\item There exists $i\geq1$ such that $\phi_{-i}\neq\psi_{-i}$. Let this $i$ be minimal.
		By definition of the total order for $\orA{k}{\lang}$, $\bigvee\tupphi \leq \bigvee\tuppsi$ if $\phi_{-i}<_\lang\psi_{-i}$.
		Assume for the sake of contradiction that $\phi_{-i}>_\lang\psi_{-i}$ (we already know they are not equal). Observe that every $j \in\{ |\tupphi|-i+1,\dots,|\tupphi| \}$ must have $m(j)>|\tuppsi|-i+1$, since otherwise $\phi_j\geq_\lang\phi_{-i}>_\lang\psi_{-i}\geq_\lang\psi_{m(j)}$, contradicting $\phi_j\sqsubseteq_\lang\psi_{m(j)}$. But this is a contradiction, since it means $m$ is not injective. Thus, $\phi_{-i}<_\lang\psi_{-i}$.
		
		\item Otherwise, $\tupphi$ is a suffix of $\tuppsi$. ($\tupphi$ cannot be longer than $\tuppsi$, since there is an injection from $\indices{\tupphi}$ to $\indices{\tuppsi}$.) This case also implies $\bigvee\tupphi \leq \bigvee\tuppsi$.
	\end{itemize}

	For $\andA{\omega}{\lang}$, given canonical $\bigwedge\tupphi \sqsubseteq \bigwedge\tuppsi$, there exists $m : \indices{\tuppsi}\to\indices{\tupphi}$ such that for all $i\in \indices{\tuppsi}$, $\phi_{m(i)} \sqsubseteq_\lang \psi_i$, and thus $\phi_{m(i)} \leq_\lang \psi_i$. Since $\tupphi$ and $\tuppsi$ are canonical, they are \emph{antichains} w.r.t $\sqsubseteq_\lang$ ordered by $\leq_\lang$. Consider the following cases:
	\begin{itemize}
		\item There exists $i\geq1$ such that $\phi_i\neq\psi_i$. Let this $i$ be minimal.
		By definition of the total order for $\andA{\omega}{\lang}$, $\bigwedge\tupphi \leq \bigwedge\tuppsi$ if $\phi_i<_\lang\psi_i$.
		Assume for the sake of contradiction that $\phi_i>_\lang\psi_i$ (we already know they are not equal). Observe that every $j\in\{1,...,i\}$ must have $m(j)<i$, since otherwise $\phi_{m(j)}\geq_\lang\phi_i>_\lang\psi_i\geq_\lang\psi_j$, contradicting $\phi_{m(j)}\sqsubseteq_\lang\psi_j$. But this means that there are distinct $1\leq j_1 < j_2 \leq i$ with $m(j_1)=m(j_2)$. Since $j_1<i$, we also know $\phi_{j_1}=\psi_{j_1}$. Thus, either $m(j_1)=m(j_2)=j_1$, which means that $\psi_{j_1}=\phi_{j_1}=\phi_{m(j_2)}\sqsubseteq_\lang\psi_{j_2}$ and $\bigwedge\tuppsi$ is not canonical; or $m(j_1)=m(j_2)\neq j_1$, which means that $\phi_{m(j_1)}\sqsubseteq_\lang\psi_{j_1}=\phi_{j_1}$ and $\bigwedge\tupphi$ is not canonical.
		Therefore, $\bigwedge\tupphi \leq \bigwedge\tuppsi$.
		
		\item Otherwise, $\tuppsi$ is a prefix of $\tupphi$. ($\tupphi$ cannot be a strict prefix of $\tuppsi$, since then there would be an element in $\tupphi$, which is also in $\tuppsi$, subsuming two distinct elements in $\tuppsi$, making $\tuppsi$ non-canonical.) This case also implies $\bigvee\tupphi \leq \bigvee\tuppsi$.
	\end{itemize}

	For $\quantA{X}{\lang}$ where $Q\in\{\bm\exists,\bm\forall\}$, given canonical $QX.\phi \sqsubseteq QX.\psi$, there exists $\pi\in\perm{X}$ such that $\phi\pi\sqsubseteq_\lang\psi$. Since $QX.\phi$ is $\quantA{X}{\lang}$-canonical,
	\[ \phi = \min{}_{\leq_\lang} \{ c_\lang(\phi\theta) \mid \theta\in\perm{X} \} \leq_\lang c_\lang(\phi\pi) \equiv_{\sqsubseteq_\lang} \phi\pi \sqsubseteq_\lang \psi, \]
	and since $\psi$ is $\lang$-canonical, $\phi\leq_\lang\psi$ and therefore $QX.\phi\leq QX.\psi$ as well. The case for $\existsforallA{X}{\lang}$ uses the two single-quantifier cases in a straightforward way.
\end{proof}

\subsection{Correctness of Set Representation}
\begin{proof}[\Cref{thm:representation}]
	Let $F \subseteq \lang$ and $R_F = \min{}_{\sqsubseteq} \{c(\phi) \mid \phi \in F\}$ its representation. Using the definition of $\min{}_{\sqsubseteq}$, and due to $\sqsubseteq$ being antisymmetric and well-founded over canonical formulas, for all $\phi\in F$ there exists $\psi\in R_F$ such that $\psi\sqsubseteq c(\phi)$. Due to $c$ being representative, $c(\phi)\equiv_\sqsubseteq\phi$.
	
	The above implies that for all $\phi\in F$ there exists $\psi\in R_F$ such that $\psi\sqsubseteq\phi$, and thus $\psi\models\phi$; from this is it easy to deduce that $\upclose{R_F} \supseteq \upclose{F}$ and $\bigwedge R_F \models \bigwedge F$. Next, by definition, any $\psi\in R_F$ has some $\phi\in F$ such that $\psi=c(\phi)$, and therefore $\psi\equiv_\sqsubseteq\phi$ and $\psi\equiv\phi$. This implies $\upclose{F} \supseteq \upclose{R_F}$ and $\bigwedge F \models \bigwedge R_F$.
	
	Taken together, we conclude that  $\bigwedge R_F \equiv \bigwedge F$ and $\upclose{R_F} = \upclose{F}$.
\end{proof}

\subsection{Correctness of Weaken}

\begin{proof}[\Cref{thm:weaken-implementation}]
	In order to prove that the implementation in \Cref{thm:weaken-implementation} satisfies the specification in \Cref{def:weaken-operator-lang}, it suffices to show that for all $\phi\in\lang$ and $\state\in\states$,
    \begin{inparaenum}[(i)]
		\item $\weaken{\lang}{\phi}{\state}$ is an antichain w.r.t $\sqsubseteq_\lang$ of canonical formulas,%
        \label{item:weaken-proof-1}
		\item all formulas in $\weaken{\lang}{\phi}{\state}$ are subsumed by $\phi$ and satisfied by $\state$, and%
        \label{item:weaken-proof-2}
		\item for all $\psi\in\lang$ with $\phi\sqsubseteq\psi$ and $\state\models\psi$, there exists $\xi\in\weaken{\lang}{\phi}{\state}$ such that $\xi\sqsubseteq\psi$.%
        \label{item:weaken-proof-3}
	\end{inparaenum}
    
    The above suffices for the proof since item (ii) shows that
    \[ \weaken{\lang}{\phi}{\state} \subseteq \upclose{\big(\min{}_{\sqsubseteq_\lang} \left\{c_\lang(\psi) \mid \psi\in\lang,  \phi\sqsubseteq\psi \text{, and } \state\models\psi\right\}\big)}, \]
    item (iii) shows that
    \[ \upclose{\weaken{\lang}{\phi}{\state}} \supseteq \min{}_{\sqsubseteq_\lang} \left\{c_\lang(\psi) \mid \psi\in\lang,  \phi\sqsubseteq\psi \text{, and } \state\models\psi\right\}, \]
    and because item (i) tells us $\weaken{\lang}{\phi}{\state}$ is an antichain of canonical formulas,
    \[ \weaken{\lang}{\phi}{\state} = \min{}_{\sqsubseteq_\lang} \left\{c_\lang(\psi) \mid \psi\in\lang,  \phi\sqsubseteq\psi \text{, and } \state\models\psi\right\}. \]
	
	As usual, we shall prove (i)--(iii)
    by induction on the recursive implementation. These clearly hold for any case where $\state\models\phi$, where the implementation always returns $\{\phi\}$. We therefore restrict our discussion to cases where $\state\not\models\phi$.
	Claim (i) can be easily verified by going case by case and using claim (i) of the induction hypothesis.
	Claim (ii) is likewise easy to prove using claim (ii) of the induction hypothesis, together with the definition of subsumption (\Cref{def:subsumption}), and the fact that a disjunction is satisfied if one of its disjuncts is satisfied, a conjunction is satsified if all of its conjuncts are satisfied, an existentially quantified formula is satisfied if there exists a state updated with an assignment to quantified variables which satisfies its body, and a universally quantified formula is satisfied if all such updated states satisfy its body.
	
	Thus, it remains to inductively prove claim (iii), which we do case by case. Keep in mind that we assume $\state\not\models\phi$.
	
	For $\lang_A$, if $\phi=\bot$, $\weaken{\lang_A}{\phi}{\state}$ contains all $\psi\in\lang_A$ with $\state\models\psi$, so (iii) holds due to reflexivity of $\sqsubseteq$. If $\phi\neq\bot$, nothing satisfying $\state$ is subsumed by it.
	
	For $\orAB{\lang_1}{\lang_2}$, assume $\phi_1\vee\phi_2 \sqsubseteq \psi_1\vee\psi_2$ and $\state\models\psi_1\vee\psi_2$.
	Either $\state\models\psi_1$, in which case there exists $\xi_1\in\weaken{\lang_1}{\phi_1}{\state}$ such that $\xi_1\sqsubseteq_{\lang_1} \psi_1$ (i.h.), and then $\xi_1\vee\phi_2 \sqsubseteq \psi_1\vee\psi_2$ and $\xi_1\vee\phi_2 \in \weaken{\orAB{\lang_1}{\lang_2}}{\phi_1\vee\phi_2}{\state}$;
	or $\state\models\psi_2$, in which case there exists $\xi_2\in\weaken{\lang_2}{\phi_2}{\state}$ such that $\xi_2\sqsubseteq_{\lang_2} \psi_2$ (i.h.), and then $\phi_1\vee\xi_2 \sqsubseteq \psi_1\vee\psi_2$ and $\phi_1\vee\xi_2 \in \weaken{\orAB{\lang_1}{\lang_2}}{\phi_1\vee\phi_2}{\state}$.
	
	For $\andAB{\lang_1}{\lang_2}$, assume $\phi_1\wedge\phi_2 \sqsubseteq \psi_1\wedge\psi_2$ and $\state\models\psi_1\wedge\psi_2$. Then $\state\models\psi_1$ and $\state\models\psi_2$. So there exist $\xi_1\in\weaken{\lang_1}{\phi_1}{\state}$ and $\xi_2\in\weaken{\lang_2}{\phi_2}{\state}$ with $\xi_1\sqsubseteq\psi_1$ and $\xi_2\sqsubseteq\psi_2$ (i.h.) (meaning $\xi_1\wedge\xi_2 \sqsubseteq \psi_1\wedge\psi_2$) such that $\xi_1\wedge\xi_2\in\weaken{\andAB{\lang_1}{\lang_2}}{\phi_1\wedge\phi_2}{\state}$.
	
	For $\orA{k}{\lang}$, assume $\bigvee\tupphi \sqsubseteq \bigvee\tuppsi$ and $\state\models\bigvee\tuppsi$. So there exists an injective $m : \indices{\tupphi} \to \indices{\tuppsi}$ such that for all $i\in\indices{\tupphi}$, $\phi_i\sqsubseteq\psi_{m(i)}$; and there exists some $j\in\indices{\tuppsi}$ such that $\state\models\psi_j$. Consider the following cases:
	\begin{itemize}
		\item There exists $i\in \indices{\tupphi}$ such that $m(i)=j$. Since $\phi_i\sqsubseteq\psi_j$ and $\state\models\psi_j$, there exists some $\xi\in\weaken{\lang}{\phi_i}{\state}$ such that $\xi\sqsubseteq\psi_j$ (i.h.), and
		\[  \bigvee\tup\xi = \bigvee \ordered{\leq_\lang}{\phi_1,\ldots,\phi_{i-1},\xi,\phi_{i+1},\ldots,\phi_{\card{\tupphi}}} \sqsubseteq \tuppsi, \]
		and $\bigvee\tup\xi \in \weaken{\orA{k}{\lang}}{\bigvee\tupphi}{\state}$, (specifically, $\bigvee\tup\xi \in W_{\card{\tupphi}}$; see \Cref{thm:weaken-implementation}).
		\item Otherwise, $m(i)\neq j$ for all $i\in \indices{\tupphi}$. Therefore, $\tupphi$ must be strictly shorter than $\tuppsi$, and therefore $\card{\tupphi}<k$. \emph{Note that this is only the case because we required $m$ to be injective when defining the subsumption relation. This is crucial for weakening disjunctions in this way.} Since $\bot_\lang \sqsubseteq \psi_j$ and $\state\models\psi_j$, there exists some $\xi\in\weaken{\lang}{\bot_\lang}{\state}$ such that $\xi\sqsubseteq\psi_j$ (i.h.), and
		\[  \bigvee\tup\xi = \bigvee \ordered{\leq_\lang}{\phi_1,\ldots,\phi_{\card{\tupphi}},\xi} \sqsubseteq \tuppsi, \]
		and $\bigvee\tup\xi \in \weaken{\orA{k}{\lang}}{\bigvee\tupphi}{\state}$, (specifically, $\bigvee\tup\xi \in W_{\card{\tupphi}+1}$; see \Cref{thm:weaken-implementation}).
	\end{itemize}

	For $\andA{\omega}{\lang}$, assume $\bigwedge\tupphi \sqsubseteq \bigwedge\tuppsi$ and $\state\models\bigwedge\tuppsi$. So there exists $m : \indices{\tuppsi} \to \indices{\tupphi}$ such that for all $i\in\indices{\tuppsi}$, $\phi_{m(i)}\sqsubseteq\psi_i$ and $\state\models\psi_i$. Therefore, for any $i\in\indices{\tuppsi}$ there exists $\xi'_i\in\weaken{\lang}{\phi_{m(i)}}{\state}$ such that $\xi'_i\sqsubseteq\phi_{m(i)}\sqsubseteq\psi_i$ (i.h.); and if we denote
	\[ \tup\xi = \ordered{\leq_\lang}{\min{}_{\sqsubseteq_\lang} \weaken{\lang}{\phi_1}{\state} \cup \cdots \cup \weaken{\lang}{\phi_\card{\tupphi}}{\state}} \]
	then each $\xi'_i$ has $j\in\indices{\tup\xi}$ such that $\xi_j\sqsubseteq\xi'_i$. Thus, any $i\in\indices{\tuppsi}$ has $j\in\indices{\tup\xi}$ such that $\xi_j\sqsubseteq \psi_i$, proving $\bigwedge\tup\xi \sqsubseteq \bigwedge\tuppsi$; and indeed, $\{\bigwedge\tup\xi\} = \weaken{\andA{\omega}{\lang}}{\bigwedge\tupphi}{\state}$.
	
	For $\existsA{X}{\lang}$, assume $\exists X.\phi\sqsubseteq \exists X.\psi$ and $((\univ,\interp),\asgn)\models\exists X.\psi$. So there exists $\pi\in\perm{X}$ such that $\phi\sqsubseteq\psi\pi$; and since $((\univ,\interp),\asgn)\models\exists X.\psi\pi$ as well, there exists $\asgnb : X \to \univ$ such that $((\univ,\interp),\stateupdate{\asgn}{\asgnb})\models\psi\pi$. Thus, there exists some $\xi\in\weaken{\lang}{\phi}{((\univ,\interp),\stateupdate{\asgn}{\asgnb})}$ such that $\xi\sqsubseteq\psi\pi$, and therefore $\exists X.\xi\sqsubseteq\exists X. \psi$; the rest of the proof follows due to canonicalization being representative.
	
	The case for $\forallA{X}{\lang}$, where we assume $\forall X.\phi\sqsubseteq \forall X.\psi$ and $\state\models\forall X.\psi$, is similar, but instead of weakening $\phi$ according to any one assignment we weaken it via
	\[
	\Omega_{\phi_0}\left(\{\state_1,\ldots,\state_n\}\right) =
	\{\phi_n \mid
	\phi_1 \in \weaken{\lang}{\phi_0}{\state_1},
	\ldots,
	\phi_n \in \weaken{\lang}{\phi_{n-1}}{\state_n}
	\}
	\]
	where we set $\phi_0 =\phi$ and the states $s_1,\dots,s_n$ to be all updates of $\state$ with assignments to $X$. The proof proceeds inductively and shows that for all $i=1,...,n$, there exist $\phi_0\in\{\phi\},\phi_1\in\weaken{\lang}{\phi_0}{s_1},...,\phi_i\in\weaken{\lang}{\phi_{i-1}}{s_i}$ subsuming $\psi\pi$, where $\pi\in\perm{X}$ is the permutation for which $\phi\sqsubseteq\psi\pi$. The induction step is similar to the case for $\existsA{X}{\lang}$, and holds since $\psi$ satisfies all updated states.
	
	The case for $\existsforallA{X}{\lang}$, where we assume $QX.\phi\sqsubseteq Q'X.\psi$ and $\state\models Q' X.\psi$, follows from the weakening procedure for $\existsA{X}{\lang}$ when $Q=Q'=\exists$, from the weakening procedure for $\forallA{X}{\lang}$ when $Q=Q'=\forall$, and handles the case $Q=\forall,Q'=\exists$ by including the result of weakening $\exists X.\phi$ as well if $Q=\forall$.
\end{proof}

\section{The $\lset{\lang}$ Data Structure}
\label{app:lset}

We introduce the $\lset{\lang}$ data structure for storing sets of canonical $\lang$-formulas (the sets are not necessarily antichains, see discussion below), and implementing the $\unsat{\cdot}{\state}$ and $\subsuming{\cdot}{\psi}$ filters.

\begin{table}
\caption{The $\lset{\lang}$ data structure. \label{table:lset}}
\centering
\begin{adjustbox}{width=\textwidth,keepaspectratio}
\begin{tblr}{
  cell{1}{1} = {r=2}{},
  cell{3}{1} = {r=2}{},
  cell{5}{1} = {r=2}{},
  cell{7}{1} = {r=3}{},  
  cell{10}{1} = {r=2}{},
  cell{9}{2} = {c=2}{},  
  cell{12}{1} = {r=2}{}, 
  cell{14}{1} = {r=2}{}, 
  cell{16}{1} = {r=2}{}, 
  vline{1,2,4} = {-}{1pt},
  vline{3} = {-}{},
  hline{1,3,5,7,10,12,14,16,18} = {-}{1pt},
  hline{2,4,6,8,9,11,13,15,17} = {2-3}{},
}
\rotatebox[origin=c]{90}{$\langa$}
&
\makecell[tl]{
\textbf{Fields:} \\
$S : \stdset{\langa}$
}
&
\makecell[tl]{
\textbf{Invariants:} \\
$\top$
}
\\
&
\makecell[tl]{
\textbf{Implementing} $\unsat{}{s}$\textbf{:} \\
$\{\phi\in S \mid \state\not\models\phi\}$    
}
&
\makecell[tl]{
\textbf{Implementing} $\subsuming{}{\psi}$\textbf{:} \\
 $\{\bot, \psi\} \cap S$
}
\\ 
\rotatebox[origin=c]{90}{$\orAB{\lang_1}{\lang_2}$}
&
\makecell[tl]{
\textbf{Fields:} \\
$S : \stdset{\orAB{\lang_1}{\lang_2}}$ \\
$L : \lset{\lang_1}$ \\ 
$M : \stdmap{\lang_1}{\lset{\lang_2}}$
}
&
\makecell[tl]{
\textbf{Invariants:} \\
$\phi_1 \in L \Leftrightarrow \phi_1 \in \dom{M} \Leftrightarrow \exists \phi_2 . \, \phi_1 \vee \phi_2 \in S$ \\
$\phi_2 \in M[\phi_1] \Leftrightarrow \phi_1 \vee \phi_2 \in S$
}
\\
&
\makecell[tl]{
\textbf{Implementing} $\unsat{}{s}$\textbf{:} \\
$\{\phi_1 \vee \phi_2 \mid \phi_1 \in \unsat{L}{s}, \phi_2 \in \unsat{M[\phi_1]}{s} \}$
}
&
\makecell[tl]{
\textbf{Implementing} $\subsuming{}{\psi_1\vee\psi_2}$\textbf{:} \\
$\{\phi_1 \vee \phi_2 \mid \phi_1 \in \subsuming{L}{\psi_1}, \phi_2 \in \subsuming{M[\phi_1]}{\psi_2} \}$
}
\\
\rotatebox[origin=c]{90}{$\andAB{\lang_1}{\lang_2}$}
&
\makecell[tl]{
\textbf{Fields:} \\
$S : \stdset{\andAB{\lang_1}{\lang_2}}$ \\
$L_1 : \lset{\lang_1}$ \\ 
$L_2 : \lset{\lang_2}$ \\ 
$M_1 : \stdmap{\lang_1}{\lset{\lang_2}}$ \\
$M_2 : \stdmap{\lang_2}{\lset{\lang_1}}$
}
&
\makecell[tl]{
\textbf{Invariants:} \\
$\phi_1 \in L_1 \Leftrightarrow \phi_1 \in \dom{M_1} \Leftrightarrow \exists \phi_2 . \, \phi_1 \wedge \phi_2 \in S$ \\
$\phi_2 \in M_1[\phi_1] \Leftrightarrow \phi_1 \wedge \phi_2 \in S$ \\
$\phi_2 \in L_2 \Leftrightarrow \phi_2 \in \dom{M_2} \Leftrightarrow \exists \phi_1 . \, \phi_1 \wedge \phi_2 \in S$ \\
$\phi_1 \in M_2[\phi_2] \Leftrightarrow \phi_1 \wedge \phi_2 \in S$
}
\\
&
\makecell[tl]{
\textbf{Implementing} $\unsat{}{s}$\textbf{:} \\
$\{\phi_1 \wedge \phi_2 \mid \phi_1 \in \unsat{L_1}{s}, \phi_2 \in M_1[\phi_1] \} \, \cup$ \\
$\{\phi_1 \wedge \phi_2 \mid \phi_2 \in \unsat{L_2}{s}, \phi_1 \in M_2[\phi_2] \}$
}
&
\makecell[tl]{
\textbf{Implementing} $\subsuming{}{\psi_1\wedge\psi_2}$\textbf{:} \\
$\{\phi_1 \wedge \phi_2 \mid \phi_1 \in \subsuming{L}{\psi_1}, \phi_2 \in \subsuming{M[\phi_1]}{\psi_2} \}$
}
\\ 
\rotatebox[origin=c]{90}{$\orA{k}{\lang}$}
&
\makecell[tl]{
\textbf{Fields:} \\
$S : \stdset{\orA{k}{\lang}}$ \\
$r : \tnode$ where $\tnode$ is \\
$~~~~(\lset{\lang},\stdmap{\lang}{\tnode})$
}
&
\makecell[tl]{
\textbf{Invariants:} \\
  $\text{collect}(r) = \{ \emptytup \} \cup
  \{\tupphi \mid \exists \tuppsi.\, (\bigvee (\tupphi + \tuppsi)) \in S 
  \}$ \\
  $(L,M) : \tnode \Rightarrow (\phi \in L \Leftrightarrow \phi \in \dom{M})$ \\
  $(L,M) : \tnode \wedge (L',M') = M[\phi] \wedge \psi \in L' \Rightarrow \phi \leq_\lang \psi$
}
\\
&
\makecell[tl]{
\textbf{Implementing} $\unsat{}{s}$\textbf{:}\\
${\{ \bigvee \tupphi \mid \tupphi \in \text{collect}_{\not\rightmodels}(r,s) \text{ and } \bigvee \tupphi \in S \}}$
}
&
\makecell[tl]{
\textbf{Implementing} $\subsuming{}{\bigvee \tuppsi}$\textbf{:} \\
$\{ \bigvee \tupphi \mid \tupphi \in \text{collect}_{\sqsubseteq}(r,\tuppsi) \text{ and } \bigvee \tupphi \in S \}$
}
\\
&
\makecell[tl]{
\textbf{Recursive functions:}\\
collect($(L,M)$) = $\{ \emptytup \} \cup \{ \seq{\phi} +\tupphi \mid \phi \in L, \tupphi \in \text{collect}(M[\phi]) \}$ \\
collect$_{\not\rightmodels}$($(L,M), s$) = $\{ \emptytup \} \cup \{ \seq{\phi} + \tupphi \mid \phi \in \unsat{L}{s}, \tupphi \in \text{collect}_{\not\rightmodels}(M[\phi], s) \}$ \\
collect$_{\sqsubseteq}$($(L,M), \tuppsi$) = $\{ \emptytup \} \cup
\{ \seq{\phi} + \tupphi \mid
i \in \indices{\tuppsi},
\phi \in \subsuming{L}{\psi_i},
\tupphi \in \text{collect}_{\sqsubseteq}(M[\phi],
\seq{ \psi_j \mid j\neq i \wedge \psi_j\geq \phi}_{j=1}^{\card{\tuppsi}})\}$
}
&
\\
\rotatebox[origin=c]{90}{$\andA{\omega}{\lang}$}
&
\makecell[tl]{
	\textbf{Fields:} \\
	$S : \stdset{\andA{\omega}{\lang}}$ \\
	$L : \lset{\lang}$
}
&
\makecell[tl]{
	\textbf{Invariants:} \\
	$\psi \in L \Leftrightarrow \exists \tupphi, i\in\indices{\tupphi}.\,
    (\bigwedge \tupphi) \in S \wedge \psi =\phi_i$
}
\\
&
\makecell[tl]{
	\textbf{Implementing} $\unsat{}{s}$: \\
	$\{ \bigwedge \tupphi\in S \mid \exists i\in \indices{\tupphi}.\, \phi_i \in \unsat{L}{s} \}$
}
&
\makecell[tl]{
	\textbf{Implementing} $\subsuming{}{\bigwedge \tuppsi}$: \\
	$\{ \bigwedge \tupphi \in S \mid \forall i\in [\tuppsi].\, \exists j\in \indices{\tupphi}.\, \phi_j \in \subsuming{L}{\psi_i} \}$
}
\\
\rotatebox[origin=c]{90}{$\existsA{X}{\lang}$}
&
\makecell[tl]{
\textbf{Fields:} \\
$S : \stdset{\existsA{X}{\lang}}$ \\
$L : \lset{\lang}$
}
&
\makecell[tl]{
\textbf{Invariants:} \\
  $\phi \in L \Leftrightarrow (\exists X. \phi) \in S$
}
\\
&
\makecell[tl]{
\textbf{Implementing} $\unsat{}{((\univ, \interp),\asgn)}$: \\
$\{ \exists X. \phi \mid \phi \in \bigcap_{\nu : X\to\univ} \unsat{L}{((\univ, \interp),\stateupdate{\asgn}{\nu})} \}$
}
&
\makecell[tl]{
\textbf{Implementing} $\subsuming{}{\exists X. \psi}$: \\
$\bigcup_{\pi \in \mathfrak{S}_X} \{ \exists X. \phi \mid \phi \in \subsuming{L}{\psi_\pi} \}$ where $\psi_\pi = c_\lang(\psi \pi)$
}
\\
\rotatebox[origin=c]{90}{$\forallA{X}{\lang}$}
&
{
\textbf{Fields:} \\
$S : \stdset{\forallA{X}{\lang}}$ \\
$L : \lset{\lang}$
}
&
{
\textbf{Invariants:} \\
  $\phi \in L \Leftrightarrow (\forall X. \phi) \in S$
}
\\
&
\makecell[tl]{
	\textbf{Implementing} $\unsat{}{((\univ, \interp),\asgn)}$: \\
	$\{ \forall X. \phi \mid \phi \in \bigcup_{\nu : X\to\univ} \unsat{L}{((\univ, \interp),\stateupdate{\asgn}{\nu}} \}$
}
&
\makecell[tl]{
	\textbf{Implementing} $\subsuming{}{\forall X. \psi}$: \\
	$\bigcup_{\pi \in \mathfrak{S}_X} \{ \forall X. \phi \mid \phi \in \subsuming{L}{\psi_\pi} \}$ where $\psi_\pi = c_\lang(\psi \pi)$
}
\\
\rotatebox[origin=c]{90}{$\existsforallA{X}{\lang}$}
&
\makecell[tl]{
\textbf{Fields:} \\
$S : \stdset{\existsforallA{X}{\lang}}$ \\
$L_\exists : \lset{\lang}$
$L_\forall : \lset{\lang}$
}
&
\makecell[tl]{
\textbf{Invariants:} \\
  $\exists X. \phi \in L_\exists \Leftrightarrow (\exists X. \phi) \in S$ \\
  $\forall X. \phi \in L_\forall \Leftrightarrow (\forall X. \phi) \in S$
}
\\
&
\makecell[tl]{
\textbf{Implementing} $\unsat{}{s}$: \\
$\unsat{L_\exists}{s} \cup \unsat{L_\forall}{s}$
}
&
\makecell[tl]{
\textbf{Implementing} $\subsuming{}{Q X. \psi}$: \\
$\begin{cases}
	\subsuming{L_\forall}{\forall X.\psi} \cup \subsuming{L_\exists}{\exists X.\psi} & \text{if } Q = \exists \\
	\subsuming{L_\forall}{\forall X.\psi} & \text{if } Q = \forall
\end{cases}$
}
\end{tblr}
\end{adjustbox}
\end{table}

We explain the implementation of $\lset{\lang}$ assuming standard data structures for sets and maps.
In each case we construct the data structure %
$\lset{\lang}$
by augmenting a standard set data structure, e.g., a hash-set, denoted $S:\stdset{\lang}$, with additional fields intended to facilitate the desired operations. 
\Cref{table:lset} lists the fields used in the $\lset{\lang}$ data structure for each language constructor, the invariants maintained,
and the implementation of $\unsat{\cdot}{\state}$ and $\subsuming{\cdot}{\psi}$.

For the base case, $\lset{\langa}$ is implemented without any auxiliary fields, using straightforward iteration over the elements of $S$ for computing $\unsat{\cdot}{\state}$ and $\subsuming{\cdot}{\psi}$.

For %
$\orAB{\lang_1}{\lang_2}$, 
the data structure has---in addition to $S$---two auxiliary data fields: $L : \lset{\lang_1}$, which holds the set of first-position disjuncts present in $S$, and a map $M:\stdmap{\lang_1}{\lset{\lang_2}}$, which maps each first-position disjunct to the set of its corresponding second-position disjuncts (in our implementation $\stdmap{\cdot}{\cdot}$ is a hash-map). 
That is, the data structure maintains the invariants that a formula $\phi_1\vee\phi_2$ is in the set $S$ iff $\phi_2\in M[\phi_1]$, and that $L$ contains the same formulas as the domain (the set of keys) of $M$, which we denote $\dom{M}$.
For $\unsat{\cdot}{\state}$, both disjuncts of a formula must not be satisfied by $\state$; therefore, this filter can be computed by using $\unsat{L}{\state}$ to get all first-position disjuncts that are not satisfied by $\state$, and for each $\phi_1$ of them computing $\unsat{M[\phi_1]}{\state}$ to get the relevant second-position disjuncts. Similarly, to get all formulas subsuming $\psi_1\vee\psi_2$, both disjuncts need to be subsumed; therefore,
$\subsuming{L}{\psi_1}$ is used to get all first-position disjuncts that subsume $\psi_1$, and for each $\phi_1$ of them $\subsuming{M[\phi_1]}{\psi_2}$ is used to get the relevant second-position disjuncts subsuming $\psi_2$.
Note that $L$ may not be an antichain, even if $S$ is. For example, if $S = \{\phi_1 \vee \phi_2 , \psi_1 \vee \psi_2\}$ with $\phi_1 \sqsubseteq_{\lang_1} \psi_1$ but $\phi_2 \not\sqsubseteq_{\lang_2} \psi_2$, then $S$ is an antichain but $L$ (which holds $\{\phi_1, \psi_1\}$) is not.

The implementation for 
$\andAB{\lang_1}{\lang_2}$ is similar to the binary disjunction case for $\subsuming{\cdot}{\psi_1\wedge\psi_2}$, since the subsumption relation is defined identically. However, $\unsat{\cdot}{\state}$ requires only one of the conjuncts to be unsatisfied by $\state$. %
We therefore hold two set-map pairs as above: $L_1:\lset{\lang_1}$, $M_1:\stdmap{\lang_1}{\lset{\lang_2}}$ and $L_2:\lset{\lang_2}$, $M_2:\stdmap{\lang_2}{\lset{\lang_1}}$. All first-position conjuncts that do not satisfy $s$ can be retrieved using $\unsat{L_1}{\state}$, and collecting for each such $\phi_1$ all $\phi_1\wedge\phi_2$ where $\phi_2\in M_1[\phi_1]$ results exactly in all formulas in the set where the first conjunct is not satisfied by $s$. In a symmetrical manner we find all formulas in the set where the second conjunct is not satisfied by $s$. As mentioned above, the $\subsuming{\cdot}{\psi_1\wedge\psi_2}$ filter is implemented identically to the $\orAB{\lang_1}{\lang_2}$ case using $L_1$, $M_1$.

The case of $\lset{\orA{k}{\lang}}$ generalizes the case of $\orAB{\lang_1}{\lang_2}$, in the sense that all disjuncts need to be unsatisfied by $s$ in $\unsat{\cdot}{\state}$; and all disjuncts need to subsume a (distinct) disjunct in $\subsuming{\cdot}{\bigvee \tup \psi}$. Therefore, these operations are implemented by computing them for all disjuncts iteratively.
We use a tree structure whose edges are labeled with $\lang$-formulas, and whose paths are understood as the sequences of disjuncts as they appear in formulas $\bigvee \tupphi \in S$.
Since we consider canonical formulas, we are ensured that disjuncts are ordered according to $\leq_{\lang}$ along paths in the tree.
Each node in the tree has an $\lset{\lang}$ that stores the $\lang$-formulas labeling its outgoing edges, which represent the next possible disjuncts in the sequence of $\lang$-formulas from the root to the node. This generalizes the $L$ field in the binary case, which maintains the possible first-position disjuncts, i.e., the next disjuncts w.r.t the empty sequence.
Using this construction, $\unsat{\cdot}{\state}$ and $\subsuming{\cdot}{\bigvee \tup \psi}$ are implemented by tree traversals, with recursive applications of $\unsat{L}{\state}$ and $\subsuming{L}{\psi_i}$ for the sets $L : \lset{\lang}$ in the nodes, which are used to retrieve the next possible disjuncts not satified by $\state$, or the next possible disjuncts subsuming some disjunct in $\bigvee\tuppsi$, respectively.
In the case of $\subsuming{\cdot}{\bigvee \tup \psi}$, since each resulting disjunct needs to subsume a distinct formula in $\tup \psi$, we remove $\psi_i$ from $\tup \psi$ during the recursive traversal 
if it caused us to go down that edge.
As an optimization, when going down an edge labeled $\phi$, we also remove from $\tuppsi$ all formulas $\psi_i<_\lang\phi$, since such formulas cannot be subsumed by formulas greater or equal to $\phi$.

The case of $\lset{\andA{\omega}{\lang}}$ is simpler. In addition to the standard $S: \stdset{\andA{\omega}{\lang}}$, we maintain a field $L: \lset{\lang}$ aggregating all conjuncts from all formulas in $S$. The formulas $\bigwedge\tupphi$ which are not satisfied by $s$ are those with some conjunct in $\unsat{L}{\state}$; and the formulas $\bigwedge\tupphi$ subsuming $\bigwedge\tuppsi$ are those where for all $i\in\indices{\tuppsi}$ there is a conjunct of $\bigwedge\tupphi$ in $\subsuming{L}{\psi_i}$.
These checks are implemented in a straightforward way. An alternative approach is to store conjunctions in a tree structure similar to the LSet implementation for $\orA{k}{\lang}$, and perform $\unsat{\cdot}{\state}$ and $\subsuming{\cdot}{\bigwedge\tuppsi}$ using appropriate traversals of the tree (analoguously to the way UBTrees \cite{ubtrees} are used for finding both \emph{sub}sets and \emph{super}sets). However, observe that the savings in the $\andA{\omega}{\lang}$ case would be much more modest. For example, the implementation of $\unsat{\cdot}{\state}$ for $\orA{k}{\lang}$ uses a recursive traversal of the tree that prunes away subtrees if the formula leading to them is satisfied by $\state$ (since any disjunction containing that formula is satisfied by $\state$). For $\andA{k}{\lang}$, discarding these paths is not possible, since it suffices for \emph{some} conjunct to be violated by $\state$ for the entire conjunction to be unsatisfied by it. Thus, $\unsat{\cdot}{\state}$ would require recursively traversing the entire tree, and our experience is that the overhead of maintaining the tree structure makes this approach inefficient overall. We note, however, that the implementation of LSet for unbounded conjunction is the least optimized component of our approach, and might be improved further in future work.

The cases of $\quantA{X}{\lang}$ for $Q\in\{\exists, \forall\}$ are each implemented using an LSet $L$ over $\lang$ to store the bodies of formulas, which is used to perform $\unsat{\cdot}{\state}$ and $\subsuming{\cdot}{QX.\psi}$ by iterating over all assignments to the variables, or over all permutations of variables, respectively, and using the corresponding operations for $\lset{\lang}$. The case of $\existsforallA{X}{\lang}$ then uses these two data structures in a straightforward way.

\Cref{table:lset} explained above relies heavily on maps that use formulas as keys. However, computing hashes of formulas can be expensive, and therefore an efficient implementation of $\lset{\lang}$ requires an additional optimization. Rather than using formulas as keys, in our implementation each $\lset{\lang}$ maintains a bidirectional mapping between formulas and integer IDs, and uses the IDs whenever possible, converting back to formulas only when producing the externally-facing output. For example, the filters applied to $L$ in the implementation of $\lset{\orAB{\lang_1}{\lang_2}}$ return a set of IDs that are then used as keys for $M$, avoiding unnecessary hash computations. (Other approaches, such as memoization of hashes or hash consing, can alternatively be used.)

\section{Language Optimizations}
\label{app:language-optimizations}

\cite{duoai} uses several techniques for reducing the size of first-order languages without losing precision. \cite[Lemma 6]{duoai}, for example, implies the equivalence
\begin{align*}
	&\tup{Q}\tup{x}.\, \phi(\tup{x}_\forall,\tup{x}_\exists) \vee (\psi_1(\tup{x}_\forall,\tup{x}_\exists)\wedge \psi_2(\tup{x}_\forall))
    \\
    &\equiv
    \Big[
    \tup{Q}\tup{x}.\, \phi(\tup{x}_\forall,\tup{x}_\exists) \vee \psi_1(\tup{x}_\forall,\tup{x}_\exists)
    \Big]\wedge \Big[
	\tup{Q}\tup{x}.\, \phi(\tup{x}_\forall,\tup{x}_\exists) \vee \psi_2(\tup{x}_\forall)
    \Big].
\end{align*}
where $\tup Q \tup{x}$ is some quantifier prefix, $\tup{x}_\forall$ are the universally quantified variables in $\tup{x}$ and $\tup{x}_\exists$ are the the existentially quantified variables in $\tup{x}$.
This allows~\cite{duoai} to significantly reduce the number of formulas considered, since any formula that can be equivalently decomposed in this way is semantically redundant throughout the search for an inductive invariant.

However, as we have confirmed with the authors of~\cite{duoai}, the proof of~\cite[Lemma 6]{duoai} contains an over-generalization, and the lemma in its general form does not hold. For example, consider the formula
\[ \exists x.\forall y.\ (x\neq y) \vee (p(y) \wedge q(x)) \]
and the structure $(\univ,\interp)$ where $\univ=\{a,b\}$ and $\interp(p) = \{a\}$, $\interp(q) = \{b\}$. It is easy to verify that $(\univ,\interp)\not\models \exists x.\forall y.\ (x\neq y) \vee (p(y) \wedge q(x))$, but
\begin{align*}
	(\univ,\interp) \models \exists x.\forall y.\ (x\neq y) \vee p(y) \quad\text{and}\quad 
	(\univ,\interp) \models \exists x.\forall y.\ (x\neq y) \vee q(x).
\end{align*}
The problem with~\cite[Lemma 6]{duoai} is that it
might not work when some of the universally quantified variables are quantified after existentially quantified ones.
Nonetheless, the reduction afforded by this kind of observation is significant, so it is desirable to amend it, which we do as follows.

\begin{lemma}
	\label{lem:decomposable}
	For any quantification sequence $\forall \tup{x}\,\tup{Q}\tup{y}$ and formulas $\phi(\tup{x},\tup{y})$, $\psi_1(\tup{x},\tup{y})$, and $\psi_2(\tup{x})$, the following semantic equivalence holds
	\begin{align*}
		&\forall \tup{x}. \tup{Q}\tup{y}.\ \phi(\tup{x},\tup{y}) \vee (\psi_1(\tup{x},\tup{y})\wedge \psi_2(\tup{x}))
        \\
        &\equiv \Big[
        \forall \tup{x}. \tup{Q}\tup{y}.\ \phi(\tup{x},\tup{y}) \vee \psi_1(\tup{x},\tup{y})
        \Big]\wedge\Big[
		\forall \tup{x}. \tup{Q}\tup{y}.\ \phi(\tup{x},\tup{y}) \vee \psi_2(\tup{x})
        \Big].
	\end{align*}
\end{lemma}

Before providing a proof for this lemma, we demonstrate how it allows us to reduce the languages we are dealing with without losing expressive power. In \Cref{sec:experiments} we choose $k$-pDNF as our quantifier-free language, specified as
\[ \orAB{\orA{n}{\cdot}}{\orA{k-1}{\andA{\omega}{\cdot}}}. \]
This permits us to select different base languages of first-order literals for the $k$-pDNF clause
($\orA{n}{\cdot}$)
and for the $k$-pDNF cubes
($\orA{k-1}{\andA{\omega}{\cdot}}$).
\Cref{lem:decomposable} shows that for a bounded language with a quantifier structure starting with $\forall\tup{x}$,
including literals containing only variables from $\tup{x}$
in the cubes does not add to the expressive power of $k$-pDNF---since the decomposition described in \Cref{lem:decomposable} is able to substitute them with a conjunction of $k$-pDNF formulas where cubes contain no such literals. Note, however, that this decomposition may replace one $k$-pDNF formula with $k$-pDNF formulas that have extra literals in their $k$-pDNF clause, so this restriction may be a tradeoff between the parameter $n$ and the size of the base language for the cubes.
However, we find that in practice, using this restriction captures the languages necessary to prove safety much more precisely.

The proof for \Cref{lem:decomposable} is presented below.

\begin{proof}[\Cref{lem:decomposable}]
    We prove an equivalent second-order formulation of the claim, where existentially quantified variables in $\tup y$ are replaced by existential quantification over fresh function variables $\tup f$, similarly to the process of Skolemization:
    \begin{align*}
        &\exists \tup{f}.\forall \tup{x},\tup{z}.\ \phi(\tup{x},\tup{f}(\tup{x},\tup{z})) \vee \big(\psi_1(\tup{x},\tup{f}(\tup{x},\tup{z})) \wedge \psi_2(\tup{x})\big)
        \\
        &\equiv \Big[
        \exists \tup{f}.\forall \tup{x},\tup{z}.\ \phi(\tup{x},\tup{f}(\tup{x},\tup{z})) \vee \psi_1(\tup{x},\tup{f}(\tup{x},\tup{z}))
        \Big] \wedge \Big[
        \exists \tup{f}.\forall \tup{x},\tup{z}.\ \phi(\tup{x},\tup{f} (\tup{x},\tup{z})) \vee \psi_2(\tup{x})
        \Big].
    \end{align*}
    Specifically, $\tup{z}$ consists of the universally quantified variables in $\tup{y}$, and for all $i\in\indices{\tup f}$, $f_i(\tup v)$ replaces the $i$-th existentially quantified variable in $\tup y$, where $\tup v$ are all the universally quantified variables in $\tup{x}, \tup{z}$ preceding that existentially quantified variable in the quantifier prefix.
    To simplify the notation we denote these function applications by $\tup f(\tup x, \tup z)$, even though each of them may only depend on \emph{some} of the variables in $\tup{z}$ (but note that all of them are dependent on $\tup{x}$).
    
    Using distributivity, the left-hand side of the above is equivalent to
	\[ \exists \tup{f}.\ \Big[ \forall \tup{x},\tup{z}.\ \phi(\tup{x},\tup{f}(\tup{x},\tup{z})) \vee \psi_1(\tup{x},\tup{f}(\tup{x},\tup{z}))\Big] \wedge 
	\Big[\forall \tup{x},\tup{z}.\ \phi(\tup{x},\tup{f}(\tup{x},\tup{z})) \vee \psi_2(\tup{x})\Big], \]
	which entails the right-hand side of the claim, since if there exists an assignment to $\tup f$ satisfying both conjuncts, for each of the conjuncts there exists an assignment satisfying it.
    
	It is left to show entailment in the other direction. Let $\struct=(\univ,\interp)$ be a structure and $\mu_1,\mu_2$ two assignments to $\tup{f}$ such that
    \begin{align*}
        (\struct,\mu_{1}) & \models \Big[\forall \tup{x},\tup{z}.\ \phi(\tup{x},\tup{f}(\tup{x},\tup{z})) \vee \psi_1(\tup{x},\tup{f}(\tup{x},\tup{z}))\Big], \text{ and}\\
        (\struct,\mu_{2}) & \models \Big[\forall \tup{x},\tup{z}.\ \phi(\tup{x},\tup{f} (\tup{x},\tup{z})) \vee \psi_2(\tup{x})\Big].
    \end{align*}
	Define the assignment $\mu$ to $\tup{f}$ which for all $\tup a\in \univ^{\card{\tup x}}$ and $\tup b\in \univ^{\card{\tup z}}$ satisfies
    \[\mu(\tup{f})(\tup{a},\tup{b})  =  
    \begin{cases}
		\mu_{1}(\tup{f})(\tup{a},\tup{b}) & \text{if $(\sigma,[\tup{x} \mapsto \tup{a}]) \models \psi_2(\tup{x})$}, \\
		\mu_{2}(\tup{f})(\tup{a},\tup{b}) & \text{otherwise}.
	\end{cases}\]
	Note that this is only possible because, as we have explained, all function applications in $\tup{f}(\tup{x}, \tup{z})$ are dependent on $\tup x$. If that were not the case, whether $(\sigma,[\tup{x} \mapsto \tup{a}]) \models \psi_2(\tup{x})$ could not be used to determine $\mu(\tup{f})(\tup{a},\tup{b})$ in this way. Similarly, we cannot condition the value of $\mu(\tup{f})(\tup{a},\tup{b})$ on $\tup{b}$ (other than in the way $\tup f(\tup{x}, \tup{z})$ depends on $\tup z$), since functions may partially depend on it.
	In part, this is why this lemma cannot be generalized to~\cite[Lemma 6]{duoai}.
	
	The assignment $\mu$ shows that $\sigma$ satisfies the left-hand of the equivalence, i.e.,
	\[(\struct, \asgn) \models
	\forall \tup{x},\tup{z}.\ \Big[ \phi(\tup{x},\tup{f}(\tup{x},\tup{z})) \vee \psi_1(\tup{x},\tup{f}(\tup{x},\tup{z}))\Big] \wedge 
	\Big[\phi(\tup{x},\tup{f}(\tup{x},\tup{z})) \vee \psi_2(\tup{x})\Big],\]
	because for all $\tup a\in \univ^{\card{\tup x}}$ and $\tup b\in \univ^{\card{\tup z}}$, if $(\sigma,\mu[\tup{x}\mapsto\tup{a},\tup{z}\mapsto\tup{b}]) \models \psi_2(\tup{x})$ then the right-hand conjunct is satisfied, and the left-hand conjunct is satisfied because
    \[ \mu(\tup{f})(\tup{a},\tup{b})=\mu_{1}(\tup{f})(\tup{a},\tup{b}). \]
    Otherwise, $(\sigma,\mu[\tup{x}\mapsto\tup{a},\tup{z}\mapsto\tup{b}]) \not\models \psi_2(\tup{x})$, which together with
    \[ \mu(\tup{f})(\tup{a},\tup{b})=\mu_{2}(\tup{f})(\tup{a},\tup{b}) \]
    implies $(\sigma,\mu[\tup{x},\mapsto\tup{a},\tup{z}\mapsto\tup{b}])\models\phi(\tup{x},\tup{f}(\tup{x},\tup{z}))$, so both conjuncts are satisfied.

    Thus, entailment holds in both directions, and the equivalence is established.
\end{proof}

\section{Least Fixpoint Computation via Symbolic Abstraction}
\label{sec:sym-abs-alg}

\Cref{alg:lfp-sym-abs} presents our implementation of the symbolic abstraction algorithm~\cite{symbolic-thakur,symabs2004} for computing the least fixpoint of the best abstract transformer of a transition system $(\init,\tr)$ in the abstract domain $\powerset{\lang}$ ordered by $\supseteq$.

\paragraph{Background and Notation.}
The transition relation $\tr$ of a transition system over signature $\signature$ is a formula over signature $\doublesig$, where   
$\signature'$ is the primed copy of $\signature$ (in which each symbol is replaced with a fresh primed duplicate).  
The set of transitions consists of  the pairs of states (structures over $\signature$) that jointly satisfy $\tr$. 
More precisely, given $(\univ, \interp),(\univ', \interp')\in\structs{\signature}$ with $\univ=\univ'$, we denote by $\seq{(\univ, \interp),(\univ', \interp')}\in\structs{\doublesig}$ the structure with the same universe $\univ=\univ'$, where $\interp$ determines the interpretation of symbols from $\signature$ and $\interp'$ determines the interpretation of symbols from $\signature'$.  The set of transitions consists of $\seq{\state, \state'}$ such that $\seq{\state, \state'} \models \tr$.
The semantics of the transition system $(\init, \tr)$ is described by a (concrete) transformer $\trans: \powerset{\states} \to \powerset{\states}$ 
defined by 
\[ \trans(S) = \{\state' \in \states \mid \state' \models \init \lor \exists \state \in S. \ \seq{\state, \state'} \models \tr\}.\]
The least fixpoint of $\trans$, denoted $\lfp \trans$, is the set of reachable states of $(\init, \tr)$.
The abstract domain $\powerset{\lang}$ induces an abstract transformer $\abs{\trans} :\powerset{\lang}\to \powerset{\lang}$ defined by $\abs{\trans} = \alpha \circ \trans \circ \gamma$, which over-approximates the semantics of $(\init,\tr)$ in the sense that $\lfp \trans \subseteq \gamma(\lfp \abs{\trans})$.

\begin{algorithm}
    \caption{Least Fixpoint Computation of $\abs{\trans}$ \label{alg:lfp-sym-abs}}
    \KwIn{Transition system $(\init, \tr)$, bounded first-order language $\lang$}
    \KwOut{$R$ such that $\up{R}= \lfp \abs{\trans}$ where %
    $\abs{\trans}$ is the best abstract transformer for $(\init,\tr)$ in the abstract domain $\powerset{\lang}$
    }

    \BlankLine

    $R$ : LSet[$\lang$] := $\{\bot_{\lang}\}$\;

    \While{
    $\exists \state'. \ \state' \models \init\wedge \neg (\bigwedge R)$ or $\exists \state, \state'.\ \seq{\state, \state'}\models \bigwedge R\wedge \tr \wedge \neg (\bigwedge R)' $
    }
    {
        \tcc{loop invariant: $\up{R} \supseteq \lfp \abs{\trans}$}
        weaken($R$, $\state'$) \tcp*{according to \Cref{alg:weaken-set}}
    }
    \Return{$R$}\;
\end{algorithm}

\paragraph{}
Given $(\init,\tr)$ and $\lang$, \Cref{alg:lfp-sym-abs} computes the representation of $\lfp \abs{\trans}$. To do so, it stores and updates a set $R$, which is the representation of the corresponding abstract element computed by symbolic abstraction. \Cref{alg:lfp-sym-abs} initializes the set $R$ to the representation of $\lang$ (the least element in the abstract domain $\powerset{\lang}$), which is $\{\bot_\lang\}$. Note that $\up{R} = \lang$. 
In each iteration, a CTI $\state'$ is found and $R$ is updated to a representation of $\up{R} \join \alpha(\{\state'\})$ by applying \emph{weaken} (\Cref{alg:weaken-set}). The algorithm therefore maintains the invariant that in each iteration, $R$ is the representation of $\alpha(S)$, where $S$ is the set of CTIs obtained so far, i.e., $\up{R} = \alpha(S)$.

A CTI is computed by finding a state $\state'$ such that either $\state' \models \init \wedge \neg (\bigwedge R)$, or there exists $\state$ such that $\seq{\state, \state'} \models \bigwedge R \wedge \tr \wedge \neg (\bigwedge R)'$, where $(\bigwedge R)'$ denotes the formula over signature $\signature'$ obtained from $(\bigwedge R)$ by replacing every symbol from $\signature$ with its primed copy.
Recall that a CTI should in fact be found for the abstract element represented by $R$, but since $\bigwedge R \equiv \bigwedge \up{R}$, it suffices to compute a CTI based on the representation $R$.

\section{Optimizing the Search for CTIs}
\label{sec:solver-optimizations}
The symbolic abstraction algorithm described in \Cref{sec:sym-abs-alg} and used in \Cref{sec:eval} requires performing SMT queries in order to check the inductiveness of a set of formulas $R$.
Since solvers often struggle when given even tens of formulas (with quantifier alternations, but still in EPR), we employ
several optimizations.
First, instead of directly checking the inductiveness of $\bigwedge R$, 
we can separately check that each formula $\phi\in R$ is relatively inductive w.r.t $\bigwedge R$.
Second, before 
checking the relative inductiveness of a formula,
we check if it is implied by formulas that were already proven relatively inductive (implication queries are in general cheaper than inductiveness queries since they involve only a single copy of the signature). 
The above can also be parallelized across formulas. Third, we use incremental SMT queries, as described in \cite{p-fol-ic3}. 
Lastly, in parallel to the SMT queries, we attempt to find CTIs via simulations from previous CTIs.

\fi

\end{document}